\RequirePackage{amsmath}
\RequirePackage{graphicx}
\RequirePackage{ulem}

\documentclass[11pt]{iopart}

\usepackage{bm,color,amsmath,txfonts}

\begin{document}

\topical{Cavity magnomechanics: from classical to quantum}

\author{Xuan Zuo, Zhi-Yuan Fan, Hang Qian}
\address{School of Physics, Zhejiang University, Hangzhou 310027, China}

\author{Ming-Song Ding}
\address{Basic Education Department, Dalian polytechnic University, Dalian 116034, China}
\ead{dingms@dlpu.edu.cn}

\author{Huatang Tan}
\address{Department of Physics, Huazhong Normal University, Wuhan 430079, China}
\ead{tht@mail.ccnu.edu.cn}

\author{Hao Xiong}
\address{School of Physics, Huazhong University of Science and Technology, Wuhan 430074, China}
\ead{haoxiong@hust.edu.cn}

\author{Jie Li}
\address{Zhejiang Province Key Laboratory of Quantum Technology and Device, School of Physics, and State Key Laboratory for Extreme Photonics and Instrumentation, Zhejiang University, Hangzhou 310027, China}
\ead{jieli007@zju.edu.cn}


\begin{abstract}
Hybrid quantum systems based on magnons in magnetic materials have made significant progress in the past decade. They are built based on the couplings of magnons with microwave photons, optical photons, vibration phonons, and superconducting qubits. In particular, the interactions among magnons, microwave cavity photons, and vibration phonons form the system of cavity magnomechanics (CMM), which lies in the interdisciplinary field of cavity QED, magnonics, quantum optics, and quantum information. Here, we review the experimental and theoretical progress of this emerging field.  We first introduce the underlying theories of the magnomechanical coupling, and then some representative classical phenomena that have been experimentally observed, including magnomechanically induced transparency, magnomechanical dynamical {backaction}, magnon-phonon cross-Kerr nonlinearity, etc. We also  discuss a number of theoretical proposals, which show the potential of the CMM system for preparing different kinds of quantum states of magnons, phonons, and photons, and hybrid systems combining magnomechanics and optomechanics and relevant quantum protocols based on them. Finally, we summarize this review and provide an outlook for the future research directions in this field.
\end{abstract}

\section{Introduction}

Quantum information science and technology, exploiting the principles of quantum mechanics, have great potential to cause revolutionary advances in fields of science and engineering in the twenty-first century. The manipulation and processing of quantum information and the realization of multifunctional quantum technologies require hybrid quantum systems (HQSs) as the implementation platform, which combines different physical components with individual strengths and complementary functionalities~\cite{Xiang13,Kurizki15,Clerk20}. 

Among all HQSs, the HQS based on collective spin excitations (magnons) in magnetic materials, e.g., yttrium iron garnet (YIG), has attracted great attention and made significant progress over the past decade; see relevant reviews~\cite{Lachance-Quirion19,YLi20,Pirro21,Awschalom21,Yuan22,Rameshti22,Chumak22,Zheng23,Zhang2023}. As one of the main advantages, the magnonic system exhibits an excellent ability to coherently interact with diverse quantum systems, including microwave photons~\cite{Huebl13,Tabuchi14,Zhang14,Goryachev14,Bai15,Zhang15}, optical photons~\cite{Osada16,Zhang16,Haigh16,Osada18}, vibration phonons~\cite{Zhang2016,Li18,Potts21,Shen22}, and superconducting qubits~\cite{Tabuchi15,Quirion17,Quirion20,DXu23}.  The above couplings form the systems of cavity electromagnonics, optomagnonics, magnomechanics, and the magnon-qubit system (following the terminologies in~\cite{Lachance-Quirion19}), respectively, and constitute the basic framework of {\it hybrid magnonics}.  In particular, the successful experimental realization of the strong coupling between microwave cavity photons and magnons in a YIG crystal~\cite{Huebl13,Tabuchi14,Zhang14,Goryachev14,Bai15,Zhang15}, as theoretically predicted in~\cite{Soykal10,Soykal10b}, has kick-started active researches in the emerging field of cavity magnonics~\cite{Rameshti22}. The fast development of the field has benefited from other distinguishing advantages of the magnonic system (based on YIG), such as a large frequency tunability, a low dissipation rate, rich nonlinearities, etc.

Up to now, although there have been exhaustive reviews on the interactions between magnons and microwave cavity photons, optical cavity photons, and superconducting qubits~\cite{Lachance-Quirion19,Yuan22,Rameshti22}, and on the configurations, circuits and materials for realizing the HQS based on magnons~\cite{YLi20,Awschalom21,Chumak22,Zhang2023}, there is a lack of a thorough review on the field of cavity magnomechanics (CMM), which studies the interactions  among magnons, microwave cavity photons, and vibration phonons~\cite{Zhang2016,Li18,Potts21,Shen22}.  The present review exactly fills this gap, which includes underlying theories of the magnomechanical coupling, up-to-date experimental observations, and abundant theoretical studies/protocols, covering both classical and quantum phenomena in {the field of CMM}.  In particular, the nonlinearity provided by the magnetostrictive interaction has been identified as an essential element for the experimental observations of magnomechanically induced transparency, magnomechanical dynamical {backaction}, magnon-phonon cross-Kerr nonlinearity and magnonic frequency combs, and a prerequisite in many theoretical proposals for preparing macroscopic quantum states of a sub-millimeter sized YIG sphere involving a large number of magnons and phonons, and nonclassical states, e.g., entangled and squeezed states, of microwave fields. All of these make the CMM system a unique platform for exploring rich physics in both classical and quantum regimes, which find broad important applications in macroscopic quantum studies, quantum information processing, quantum technologies, quantum memories, quantum metrology, spectroscopy, nonlinear researches, etc.

The review is organized as follows. Section~\ref{magnomechanicaltheory} describes the fundamental theory of the magnomechanical coupling and its fully quantized interaction Hamiltonian, which lays the theoretical foundation for various applications using the magnomechanical systems. Section~\ref{classCMM} introduces a series of representative classical phenomena in {the field of CMM}, of which most have been experimentally demonstrated at room temperature. Section~\ref{quCMM} discusses a number of theoretical proposals for preparing different kinds of quantum states in the system, e.g., entangled states, squeezed states and quantum steering of magnons, phonons and photons, fully exhibiting its potential as a HQS. Section~\ref{Optomag} introduces hybrid systems combining both magnomechanics and optomechanics and relevant quantum protocols based on them.  Finally, Section~\ref{conc} summarizes this review, outlines important future research directions in {the field of CMM by applying the concepts of non-Hermiticity, $\mathcal{PT}$ symmetry, and exceptional points to the CMM system,}
and briefly introduces other magnomechanical platforms beyond the frameworks set by Refs.~\cite{Lachance-Quirion19,Yuan22,Rameshti22}.

\section{Fundamental theory of the magnomechanical coupling}
\label{magnomechanicaltheory}

The magnomechanical coupling describes the interaction between magnetization and elastic strain of a magnetic material. Depending on the distance between magnetic atoms (or ions), there are three kinds of interactions: the spin-orbital interaction, the exchange interaction, and the magnetic dipole-dipole interaction \cite{Gurevich96}. For a cubic crystal, the magnomechanical coupling is described by the magnetoelastic energy density, given by \cite{Kittel49,Kittel58}
\begin{align}
	\eqalign{ f_{\rm me}=\frac{B_1}{M^2_\mathrm{S}}\left ( M_x^2\epsilon_{xx}+M_y^2\epsilon_{yy}+M_z^2\epsilon_{zz} \right ) +\frac{2 B_2}{M^2_\mathrm{S}}\left( M_x M_y\epsilon_{xy}+M_x M_z\epsilon_{xz}+M_y M_z\epsilon_{yz} \right),}
	\label{density}
\end{align}
where $B_1$ and $B_2$ are the magnetoelastic coupling coefficients, $M_{\mathrm{S}}$ is the saturation magnetization and $M_{x,y,z}$ are the magnetization components. $\epsilon_{ij}=\frac{1}{2}\left ( \partial u_{i}/\partial {l_j}+ \partial u_{j}/\partial {l_i} \right )$ is the strain tensor of the magnetic crystal, with coordinate indices $i,j\in \{x,y,z\}$, and $u_{x,y,z}$ are the components of the displacement vector $\vec{u}$.

A magnon is viewed as a quantized spin wave. According to the Holstein-Primakoff {approximation}~\cite{HP}, the magnetization can be quantized via
\begin{align}\label{MxMy}
	\eqalign{
		M_x=\sqrt{\frac{\hbar \gamma M_\mathrm{S}}{2V}}(m+m^\dagger ), \ \ 	M_y=i\sqrt{\frac{\hbar \gamma M_\mathrm{S}}{2V}}(m^\dagger -m ),}
\end{align}
and 
\begin{align} \label{Mzz}
	M_z=\left(M^2_\mathrm{S}-M_x^2-M_y^2\right)^{\frac{1}{2}}\simeq M_\mathrm{S}-\frac{\hbar\gamma}{V}m^\dagger m,
\end{align}
where $m$ ($m^\dagger$) is the annihilation (creation) operator of the magnon mode, $V$ is the volume of the magnetic crystal, and $\gamma$ is the gyromagnetic ratio. {Note that equation~\eqref{Mzz} is obtained by taking the first-order Taylor series expansion of $\sqrt{1-(M_x^2+M_y^2)/M_\mathrm{S}^2}$.}
 Substituting equations~\eqref{MxMy} and \eqref{Mzz} into the magnetoelastic energy density $f_{\rm me}$ and integrating over the whole volume of the crystal, the semiclassical magnetoelastic Hamiltonian can be obtained as the sum of 
\begin{align}
	\label{Hme1}
	\eqalign{
		H_1\simeq\ &\frac{B_1}{M_\mathrm{S}}\frac{\hbar\gamma}{V}m^\dagger m\int dl^3 \left( \epsilon_{xx}+\epsilon_{yy}-2\epsilon_{zz} \right) + \frac{B_1}{M_\mathrm{S}}\frac{\hbar\gamma}{2V} \left(m^2+m^{\dagger 2} \right) \int dl^3 \left(\epsilon_{xx}-\epsilon_{yy} \right)\\
		&+\frac{B_1}{M_\mathrm{S}^2}\frac{\hbar^2\gamma^2}{V^2}m^\dagger m m^\dagger m \int dl^3\epsilon_{zz}\ ,
	}
\end{align}
and	
\begin{align}\label{Hme2}
	\eqalign{
		H_2 {\simeq} & \ 
		{-}i\frac{B_2}{M_\mathrm{S}}\frac{\hbar\gamma}{V} \left( m^2{-}m^{\dagger 2} \right) \! \int \! dl^3 \epsilon_{xy}
		+\frac{2B_2}{M_\mathrm{S}^2} \! \sqrt{\frac{\hbar\gamma M_\mathrm{S}}{2V}}\left ( M_\mathrm{S}{-}\frac{\hbar\gamma}{V}m^\dagger m \right )\left [ m \! \int\! dl^3 \left(\epsilon_{xz} {-} i\epsilon_{yz} \right)+\mathrm{H.c.} \right ].
	}
\end{align}
The magnetoelastic Hamiltonian, $H_{\mathrm{me}}=H_1+H_2$, implies different types of magnon-phonon interactions depending on the relation of their frequencies. For specific magnon and phonon frequencies, a certain type of coupling can be dominant in the magnetoelastic coupling, while other coupling mechanisms are negligible. This can be more clearly seen in the fully-quantized interaction Hamiltonian by further quantizing the strain displacement~\cite{Fan23}.

The magnetoelastic displacement can be decomposed and expressed as the superposition form of
\begin{align}
	\vec{u}(x,y,z)=\sum_{n,m,k} d^{(n,m,k)}\vec{\chi}^{(n,m,k)}(x,y,z),
\end{align}
where $\vec{\chi}^{(n,m,k)}(x,y,z)$ represents the dimensionless displacement eigenmode and $d^{(n,m,k)}$ is the corresponding amplitude with the mode indices ($n,m,k$). The displacement amplitude $d^{(n,m,k)}$ can be quantized as
\begin{align}
	d^{(n,m,k)}=d_{\mathrm{zpm}}^{(n,m,k)}\left(b_{n,m,k}+b^\dagger_{n,m,k}\right),
\end{align}
where $d_{\mathrm{zpm}}^{(n,m,k)}$ denotes the amplitude of the zero-point motion, and $b_{n,m,k}$ and $b^\dagger_{n,m,k}$ are the annihilation and creation operators of the corresponding phonon mode.

The fully-quantized magnetoelastic Hamiltonian can be derived by quantizing the strain displacement in the semiclassical Hamiltonians of equations~\eqref{Hme1} and \eqref{Hme2}~\cite{Fan23}, which accounts for different types of magnon-phonon interactions dependent on their frequencies. Specifically, for the case of the phonon frequencies being much lower than the magnon frequency $\omega_b^{(n,m,k)}\ll\omega_m$ (the case of a large-size magnetic crystal), by neglecting the nonresonant fast-oscillating terms, one obtains the dominant dispersive-type interaction
\begin{align}
	H_{\mathrm{me}}\simeq\sum_{n,m,k}\hbar g_{\mathrm{disp}}^{(n,m,k)}m^\dagger m \left(b_{n,m,k}+b^\dagger _{n,m,k} \right),
\end{align}
where the dispersive coupling strength
\begin{align}
	g_{\mathrm{disp}}^{(n,m,k)}=\frac{B_1}{M_\mathrm{S} }\frac{\gamma}{V}\int dl^3\ d_{\mathrm{zpm}}^{(n,m,k)}\left ( \frac{\partial \chi_x^{(n,m,k)}}{\partial x} +\frac{\partial \chi_y^{(n,m,k)}}{\partial y}-2\frac{\partial \chi_z^{(n,m,k)}}{\partial z} \right ).  
\end{align}
Note that the above Hamiltonian is derived under the condition of low-lying magnon excitations, where the second-order term of the magnon excitation $m^\dagger m$ is neglected~\cite{Fan23}.  Such a dispersive coupling is the dominant magnon-phonon interaction for sub-millimeter sized YIG spheres~\cite{Zhang2016,Li18,Potts21,Shen22}, which provides essential nonlinearity that is exploited in many proposals for preparing quantum states, as will be discussed in Section~\ref{quCMM}.

For the case where the phonon mode is nearly resonant with the magnon mode, $\omega_b^{(n,m,k)}\simeq\omega_m$, after neglecting the nonresonant fast-oscillating terms, one gets the following dominant linear interaction
\begin{align}
	H_{\mathrm{me}}\simeq\sum_{n,m,k}\hbar \left ( g_{\mathrm{lin}}^{(n,m,k)}m b^\dagger_{n,m,k}+\mathrm{H.c.}  \right ),
\end{align}
where the coupling rate
\begin{equation}
	\eqalign{g_{\mathrm{lin}}^{(n,m,k)}=&\frac{B_2}{M_{\mathrm{S}}}\sqrt{\frac{\gamma M_{\mathrm{S}}}{2\hbar V}} \left [ \int dl^3\ d_{\mathrm{zpm}}^{(n,m,k)}\left ( \frac{\partial \chi_x^{(n,m,k)}}{\partial z} +\frac{\partial \chi_z^{(n,m,k)}}{\partial x}\right ) \right.\\
	&\left. -i\int dl^3\ d_{\mathrm{zpm}}^{(n,m,k)}\left ( \frac{\partial \chi_y^{(n,m,k)}}{\partial z} +\frac{\partial \chi_z^{(n,m,k)}}{\partial y}\right )  \right ].}
\end{equation}
Similarly, the higher-order term has been neglected under the low-lying excitations. This magnon-phonon linear coupling is typically the case for high-frequency phonons, e.g., in  the gigahertz range. In this regime, the magnon-phonon strong coupling can be achieved forming magnon-phonon polarons~\cite{Godejohann20}. Though strong, such a beam-splitter coupling itself can only realize state-swap interaction between magnons and phonons, but cannot create entanglement and squeezing, as will be discussed in Section~\ref{quCMM}.

Apart from the aforementioned dispersive and linear interactions, a magnon parametric amplification (PA) interaction becomes dominant when the phonon frequency is about twice of the magnon frequency, $\omega_b^{(n,m,k)}\simeq2\omega_m$.  This leads to the following interaction
\begin{align}
	H_{\mathrm{me}}\simeq\sum_{n,m,k}\hbar \left ( g_{\mathrm{PA}}^{(n,m,k)}m^2 b^\dagger_{n,m,k}+\mathrm{H.c.}  \right ),
\end{align} 
with the coupling rate
\begin{align}
	g_{\mathrm{PA}}^{(n,m,k)}=\frac{1}{M_{\mathrm{S}}}\frac{\gamma}{2V} \int dl^3\ d_{\mathrm{zpm}}^{(n,m,k)}  \left [ B_1 \left ( \frac{\partial \chi_x^{(n,m,k)}}{\partial x} -\frac{\partial \chi_y^{(n,m,k)}}{\partial y}\right ) -iB_2\left ( \frac{\partial \chi_x^{(n,m,k)}}{\partial y} +\frac{\partial \chi_y^{(n,m,k)}}{\partial x}\right )  \right ].
\end{align}
Such magnon PA can be used to generate entangled pairs of magnons, in analogy to the generation of entangled photon pairs by optical parametric down-conversion.

To sum up, by fully quantizing the magnetoelastic Hamiltonian, different {\it effective} magnomechanical Hamiltonians are derived for different {relations between} the magnon and phonon frequencies: specifically, $i$) the dispersive interaction for the phonon frequency being much lower than the magnon frequency, $\omega_b^{(n,m,k)}\ll\omega_m$; $ii$) the linear (beam-splitter) interaction for nearly resonant magnons and phonons, $\omega_b^{(n,m,k)}\simeq\omega_m$; and $iii$) the magnon PA interaction when $\omega_b^{(n,m,k)}\simeq 2\omega_m$. Note that all the effective interactions are obtained by neglecting nonresonant fast-oscillating terms, which are good approximations when the magnon frequency and (or) the phonon frequency are (is) much larger than their coupling strength and dissipation rates~\cite{Fan23}.

\section{Classical cavity magnomechanics}
\label{classCMM}

\begin{figure}[t]
	\includegraphics[width=420pt]{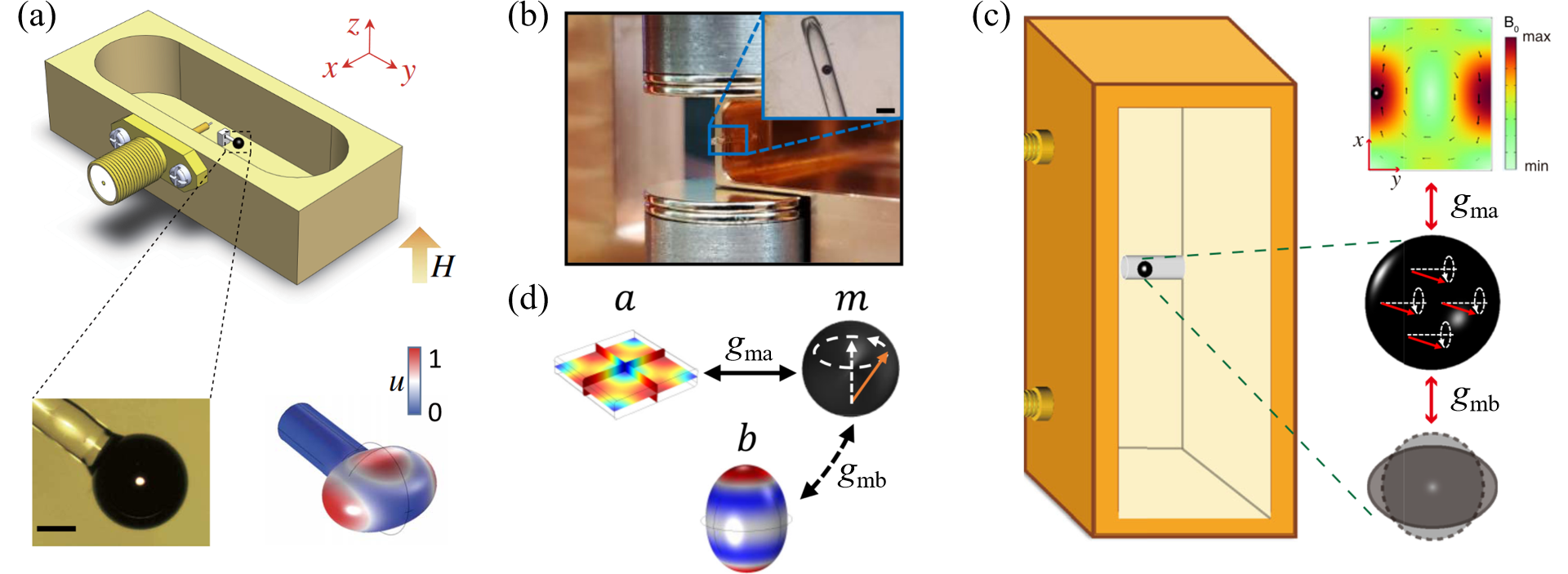}
	\centering
	\caption{{The CMM experimental device consists of a three-dimensional microwave cavity made from oxygen-free copper and a YIG sphere supporting both a magnon mode and a mechanical vibration mode. The YIG sphere is placed near the maximum magnetic field of the cavity mode and can be either glued to the end of a silica fiber~\cite{Zhang2016} (a) or placed in a horizontal glass capillary~\cite{Potts21, Shen22} (b)-(c). A uniform external magnetic field, provided by a set of permanent magnets (b), is applied to bias the YIG sphere for the magnon-photon coupling. (d) Interactions between the different modes. The magnon mode couples to the cavity magnetic field via the magnetic-dipole interaction with the coupling strength $g_{\rm ma}$, and to the mechanical vibration mode via the dispersive magnetostrictive interaction with the bare magnomechanical coupling strength $g_{\rm mb}$. Figures (a)-(d) are adapted from Refs.~\cite{Zhang2016,Potts21,Shen22}.}}
\label{setup}
\end{figure}

The exploration of the CMM has started from the experimental studies on classical phenomena, and so far, all the CMM experiments have been operated at room temperature. The typical CMM experimental device is shown in Figure~\ref{setup}, which consists of a three-dimensional microwave cavity machined from oxygen-free copper and a highly polished single-crystal YIG sphere supporting both a magnon mode and a mechanical vibration mode. The YIG sphere with the diameter of hundreds of micrometers is placed near the maximum microwave magnetic field of the cavity mode. The YIG sphere can be either glued to the end of a silica fiber~\cite{Zhang2016}, or placed (free to move) in a horizontal glass capillary to reduce the mechanical damping~\cite{Potts21,Shen22}. A uniform external magnetic field is applied to bias the YIG sphere for the magnon-photon coupling.

The Hamiltonian of the CMM system reads~\cite{Zhang2016,Li18,Potts21,Shen22}
\begin{equation}\label{HHH}
H/\hbar \!=\! \omega_{a} a^\dagger a + \omega_{m} m^\dagger m + \frac{\omega_b}{2}\! \left(q^2 {+} p^2 \right) + g_{\rm ma}\! \left(a m^\dagger {+} a^\dagger m \right) + g_{\rm mb} m^\dagger m q + H_{\rm dri}/\hbar , \,\,\,
\end{equation}
where $a$ ($a^\dagger$) and $m$ ($m^\dagger$) ([$j$, $j^\dagger$] = 1, $j$ = $a$, $m$) are the annihilation (creation) operators of the cavity and magnon modes, respectively, $q=(b+b^\dagger)/\sqrt{2}$ and $p=i(b^\dagger-b)/\sqrt{2}$ are the dimensionless position and momentum of the vibration mode (modeled as a mechanical oscillator and [$q$, $p$] = $i$), and $\omega_k$ ($k=a,m,b$) are their resonance frequencies. In typical CMM experiments, $\omega_a$ and $\omega_m$ are nearly resonant and around $\sim$10 GHz~\cite{Zhang2016,Potts21,Shen22,Shen23}. The magnon frequency can be tuned by varying the external bias magnetic field $H$ via $\omega_m = \gamma H$, where the gyromagnetic ratio $\gamma/2\pi =28$ GHz/T.  The magnon mode couples to the cavity magnetic field via the magnetic-dipole interaction with the coupling strength $g_{\rm ma}$, which can be (much) stronger than the cavity and magnon dissipation rates $\kappa_a$ and $\kappa_m$, leading to the hybridization of the two modes forming two polaritons~\cite{Huebl13,Tabuchi14,Zhang14,Goryachev14,Bai15,Zhang15}. 
The magnon mode further couples to a mechanical vibration mode via magnetostriction.  For a YIG sphere with the diameter in the range of $10^2$-$10^3$ $\mu$m, the mechanical frequency is in the megahertz range~\cite{Zhang2016,Potts21,Shen22,Shen23}, which is much lower than the magnon frequency in the gigahertz range, thus promising a dominant dispersive magnomechanical coupling~\cite{Zhang2016,Fan23,Ballestero20} (Section~\ref{magnomechanicaltheory}). The bare magnomechanical coupling $g_{\rm mb}$ is typically weak, but the effective coupling can be significantly enhanced by driving either the magnon or the cavity with a strong microwave field. For driving the magnon, e.g., via a loop antenna~\cite{Shen23}, $ H_{\rm dri}/\hbar = i\Omega \left(m^\dagger e^{-i\omega_0t}-m e^{i\omega_0t} \right)$, where the Rabi frequency $\Omega=\frac{\sqrt{5}}{4} \gamma \sqrt N B_0$~\cite{Li18} denotes the coupling strength between the magnon and the drive magnetic field with frequency $\omega_0$ and amplitude $B_0$. $N = \rho V$ is the total number of spins with $\rho = 4.22 \times 10^{27}$ $\rm{m^{-3}}$ being the spin density of the YIG and $V$ being the volume of the YIG crystal. For driving the cavity, $ H_{\rm dri}/\hbar = i E \left(a^\dagger e^{-i\omega_{d}t}-a e^{i\omega_{d}t} \right)$, where $E$ denotes the cavity-drive coupling strength and $\omega_{d}$ is the frequency of the drive field.

In what follows, we introduce a series of classical phenomena that have been either experimentally demonstrated or theoretically studied, including magnomechanically induced transparency and absorption, magnomechanical dynamical {backaction}, magnon-phonon cross-Kerr nonlinearity and mechanical bistability, etc.

\subsection{Magnomechanically induced transparency and absorption}

The radiation-pressure-like dispersive magnomechanical interaction predicts the magnomechanically induced transparency (MMIT) and absorption (MMIA)~\cite{Zhang2016}, in analogy to the optomechanically induced transparency (OMIT) and absorption (OMIA)~\cite{Weis10,Agarwal10,Xiong18}. It describes an interference phenomenon that the transmission/reflection of the weak probe signal is controlled by a strong drive field when the resonance condition is met. In a typical measurement setup of the MMIT (MMIA) \cite{Zhang2016}, the cavity is driven by both a strong microwave field with frequency $\omega_d$ and amplitude $\varepsilon_{d}$ and a weak probe signal with frequency $\omega_s$ and amplitude $\varepsilon_s$.  The dynamics of the system is described by the following semiclassical Heisenberg equations:
\begin{equation}\label{mmit2}
\eqalign{
		\dot a =& ( - i{\omega _a} - {\kappa _{a}})a - i{g_{{\rm{ma}}}}m - i\sqrt {2{\kappa _{e}}} {\varepsilon _{d}}{e^{ - i{\omega _{d}}t}} - i\sqrt {2{\kappa _{e}}} {\varepsilon _{s}}{e^{ - i{\omega _{s}}t}},  \\
		\dot{m}=&( - i{\omega _m} - {\kappa _m})m - i{g_{{\rm{ma}}}}a - i{g_{{\rm{mb}}}}({b^{\rm{*}}} + b)m, \\
		\dot{b}=&( - i{\omega _b} - {\gamma _b})b - i{g_{{\rm{mb}}}}{m^{\rm{*}}}m,}
\end{equation}
where ${{\kappa _{e}}}$ is the external decay rate of the cavity, and $\gamma _b$ is the mechanical damping rate.  Note that the noise terms are removed and the operators are replaced with their expectation values. Equivalently, the position and momentum $(q,p)$ are replaced with $(b,b^*)$ to describe the mechanical mode.

\begin{figure}[t]
	\includegraphics[width=430pt]{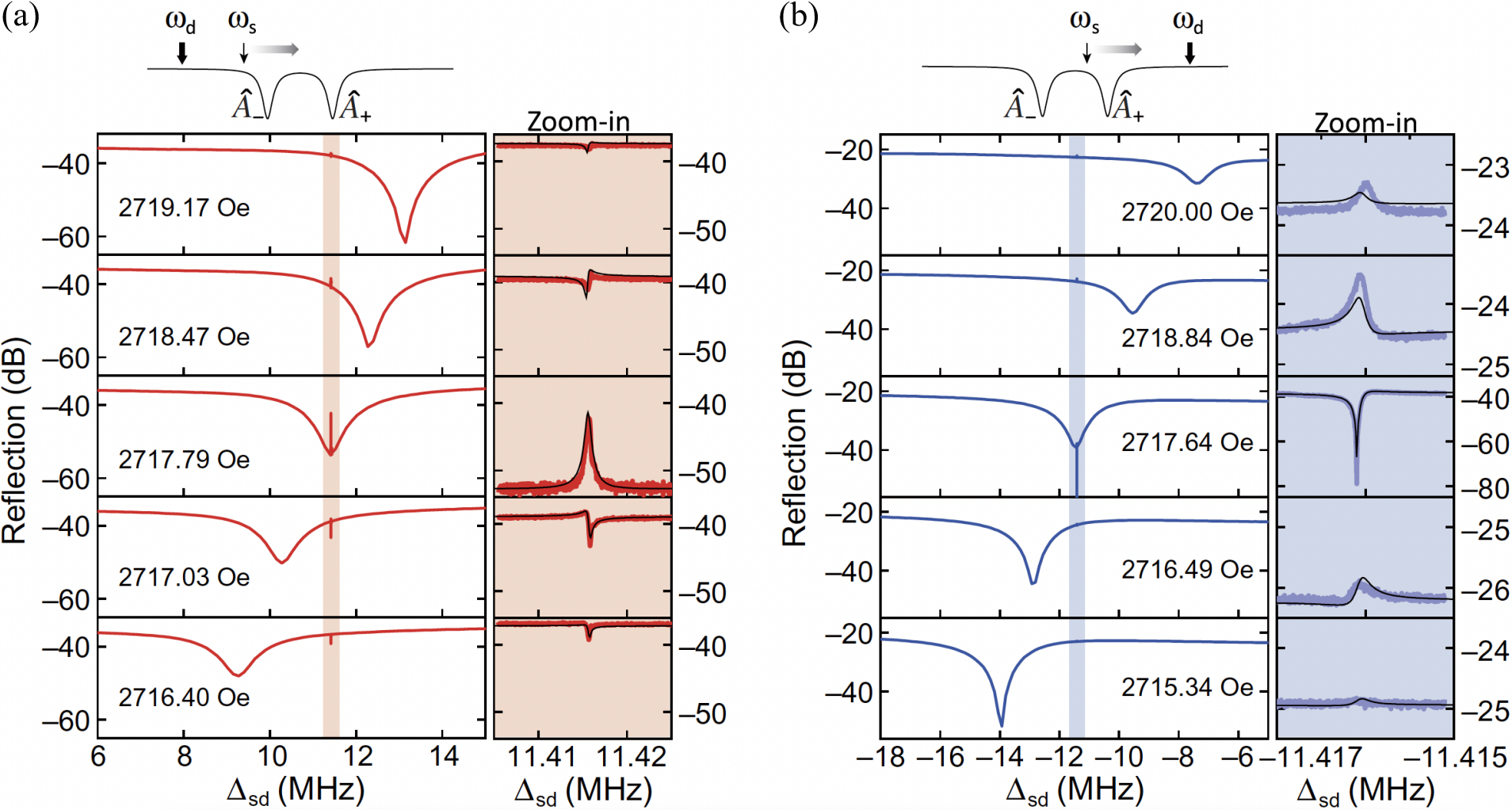}
	\centering
	\caption{Measured reflection spectra as a function of the probe-drive detuning $\Delta_{\mathrm{sd}}$ (with a fixed driving frequency $\omega_d$) for a series of different bias magnetic fields, for (a) a red-detuned drive $\omega_d <\omega_-$; (b) a blue-detuned drive $\omega_d >\omega_+$. Zoom-in shows detailed spectra of the magnomechanically induced resonances. The figures are adapted from Ref.~\cite{Zhang2016}. 
}
\label{figmmit}
\end{figure}

The Heisenberg equations (\ref{mmit2}) are nonlinear due to the dispersive magnomechanical coupling. They can be solved by linearizing the dynamics of the system around the steady-state values by writing $O=\bar{O}+\delta O$ ($O=a,m,b$), where (${\bar a}, {\bar m}, {\bar b}$) are the large mean values and ($\delta a$, $\delta m$, $\delta b$) are the small perturbations. Because the probe signal is much weaker than the drive field, the probe signal can be neglected in deriving the solutions of the mean fields, which are given by
 \begin{gather}
\bar a = \frac{{i{g_{{\rm{ma}}}}\bar m + i\sqrt {2{\kappa _{e}}} {\varepsilon _{d}}}}{{i{\omega _{d}} - i{\omega _a} - {\kappa _{a}}}},\qquad\qquad \bar b = \frac{{i{g_{{\rm{mb}}}}{{\left| {\bar m} \right|}^2}}}{{ - i{\omega _b} - {\gamma _b}}},\nonumber\\
\frac{{2ig_{{\rm{mb}}}^2{\omega _b}}}{{\gamma _b^2 + \omega _b^2}}{\left| {\bar m} \right|^2}\bar m + \left( {i{\omega _{d}} - i{\omega _m} - {\kappa _m} + \frac{{g_{{\rm{ma}}}^2}}{{i{\omega _{d}} - i{\omega _a} - {\kappa _{a}}}}} \right)\bar m + \frac{{{g_{{\rm{ma}}}}\sqrt {2{\kappa _{e}}} {\varepsilon _{d}}}}{{i{\omega _{d}} - i{\omega _a} - {\kappa _{a}}}} = 0.
\end{gather}
By omitting high-order perturbation terms, one obtains the linearized equations for the perturbations ($\delta a$, $\delta m$, $\delta b$), which, in the frame rotating at the drive frequency, are
\begin{equation}
\eqalign{
 \frac{{\rm{d}}}{{{\rm{d}}t}}\delta a = (i{\omega _{d}} - i{\omega _a} - {\kappa _{a}})\delta a - i{g_{{\rm{ma}}}}\delta m - {{i}}\sqrt {2{\kappa _{e}}} {\varepsilon _{s}}{e^{ - i({\omega _{s}} - {\omega _{d}})t}}, \\
 \frac{{\rm{d}}}{{{\rm{d}}t}}\delta m = \left[ {i{\omega _{d}} - i{\omega _m} - {\kappa _m} - i{g_{{\rm{mb}}}}({{\bar b}^{\rm{*}}} + \bar b)} \right]\delta m - i{g_{{\rm{ma}}}}\delta a - i{g_{{\rm{mb}}}}\bar m(\delta {b^{\rm{*}}} + \delta b), \\
 \frac{{\rm{d}}}{{{\rm{d}}t}}\delta b = ( - i{\omega _b} - {\gamma _b})\delta b - i{g_{{\rm{mb}}}}(\delta {m^{\rm{*}}}\bar m + {{\bar m}^{\rm{*}}}\delta m).
 }
\end{equation}
The above equations can be conveniently solved in the frequency domain by taking the Fourier transform of each equation.
{Having achieved the solution of $\delta a(\omega)$, one then obtains the cavity output field $\delta a_{\rm out}(\omega)$ using the input-output relation ${\delta a_{{\rm{out}}}(\omega)} = {\varepsilon _{s}} + i\sqrt {2{\kappa _{e}}} \delta a(\omega)$~\cite{Zhang2016}, and thereby the reflection spectrum of the probe signal, i.e., $r(\omega)\equiv {\delta a_{{\rm{out}}}(\omega)}/{\varepsilon _{s}}$.}
The measured reflection spectra as a function of the probe-drive detuning $\Delta_{\mathrm{sd}}={\omega _{s}} - {\omega _{d}}$ for a red (blue)-detuned  drive field is shown in Figure~\ref{figmmit}(a) (Figure~\ref{figmmit}(b)). The sharp Lorentzian-shaped peak (dip) at $\Delta_{\mathrm{sd}}= \omega_b$ ($\Delta_{\mathrm{sd}}= -\omega_b$) in the spectra is evidence of the magnomechanical interaction, i.e., the MMIT (MMIA) in Figure \ref{figmmit}(a) (Figure \ref{figmmit}(b))~\cite{Zhang2016}. Note that in each driving case, the driving frequency $\omega _{d}$ is fixed and the mechanical frequency $\omega_b/2\pi \approx 11.42$ MHz in the experiment~\cite{Zhang2016}.

The strong coupling between the cavity and magnon modes leads to two hybridized polariton modes $A_{\pm}$, formed via $A_+=a \cos\theta + m\sin\theta$ and $A_-=- a\sin\theta + m\cos\theta$, with $ \theta =\frac{1}{2} \arctan \frac{2g_{\rm ma}}{\omega_a-\omega_m}$. It indicates that the weight of the cavity (magnon) component in each polariton can be varied by altering $\theta$, realized by changing the coupling $g_{\rm ma}$ and/or the detuning $\omega_a-\omega_m$. Since the cavity component has no coupling with the mechanical mode, the interaction between each polariton and the mechanics is fully determined by the interaction between the magnon component and the mechanics, i.e., the polariton-mechanics coupling is also dispersive~\cite{Zuo23,Shen23}. In the spectra of Figure~\ref{figmmit}(a) (Figure~\ref{figmmit}(b)), the broad Lorentzian resonance dip corresponds to the polariton mode $A_-$ ($A_+$), whose frequency $\omega_-$ ($\omega_+$) can be tuned by changing the magnon frequency realized by varying the bias magnetic field. When the polariton $A_-$ ($A_+$) is tuned to be resonant with the anti-Stokes (Stokes) mechanical sideband of the drive field, i.e., at $\Delta_{\mathrm{sd}}= \omega_b$ ($\Delta_{\mathrm{sd}}= -\omega_b$) in Figure~\ref{figmmit}(a) (Figure~\ref{figmmit}(b)), the MMIT (MMIA) is most prominent, due to the strong interference between the two resonances.

Similarly as the OMIT (OMIA), the MMIT (MMIA) can also cause characteristic changes in dispersion in the range of the transparency (absorption) window, which can be used to manipulate the group velocity of the probe signal. Slow light effects are predicted to occur in the CMM system \cite{Kong19}. Long-lived slow light (group delay of the order of millisecond) can be achieved and the group delay can be continuously adjusted by tuning the bias magnetic field \cite{Kong19}. Multi-window MMIT, Fano resonances, and slow-to-fast light conversion have been studied in a two-YIG-spheres CMM system \cite{Ullah20}, where the group delay of the probe signal can be greatly enhanced compared to the single-YIG-sphere case. It is also suggested that magnon squeezing plays an active role in enhancing and controlling the group delay of the probe field~\cite{Lu23} and phase control of the slow light effect is possible \cite{Li20}.

\subsection{Magnomechanical dynamical backaction: mechanical cooling and amplification, and magnonic spring effect}
\label{backac}

The dispersive magnomechanical interaction also promises dynamical backaction effects on the mechanical mode (similar as those in optomechanics~\cite{Aspelmeyer14}), such as mechanical cooling and amplification, and the ``magnonic spring" effect, i.e., the mechanical frequency modified by the magnetostrictive interaction~\cite{Potts21}.  The characterization of these effects requires the measurements of the frequency and linewidth of the mechanical mode.  The mechanical frequency and linewidth, as well as the bare magnomechanical coupling $g_{\rm mb}$, {can be extracted by fitting the MMIT signal~\cite{Zhang2016,Shen22}, or adopting a microwave homodyne detection scheme, which consists of a microwave generator as the local oscillator, an IQ mixer, and an analog-to-digital converter~\cite{Potts21}. They play a similar role as in the optical homodyne detection of the mechanical state in optomechanics~\cite{deJong22}.}

To enhance the magnomechanical coupling, the system is operated to satisfy the triple-resonance condition~\cite{Zhang2016}, where the splitting of the two hybrid modes (i.e., two polaritons) in the spectrum matches the mechanical frequency, $\omega_+-\omega_- \equiv \sqrt{4g_{\rm ma}^2 + \Delta_{\rm am}^2} =\omega_b$, with $\Delta_{\rm am}\equiv \omega_a-\omega_m$. In this case, both the drive field and the scattered mechanical sideband can be on resonance with the two polaritons, respectively, leading to drastically enhanced magnomechanical coupling.  In the experiment~\cite{Potts21}, this is implemented by setting the normal mode splitting (NMS) for the resonant case ($\Delta_{\rm am}=0$) to be slightly smaller than the mechanical frequency, i.e., $2g_{\rm ma} < \omega_b$.  Then by slightly detuning the magnon frequency from the cavity frequency ($\Delta_{\rm am}\ne 0$), the NMS can be tuned to exactly match the mechanical frequency. When the triple-resonance condition is met, one can choose to resonantly drive the lower (higher)-frequency polariton, such that the anti-Stokes (Stokes) mechanical sideband resonates with the higher (lower)-frequency polariton, resulting in the cooling (amplification) of the mechanical motion~\cite{Potts21}. This corresponds to an increased (a reduced) effective mechanical damping rate $\gamma_b^{\rm eff} = \gamma_b + \delta\gamma_b$, with $\delta\gamma_b >0$ ($\delta\gamma_b <0$) being the magnomechanically induced additional damping rate, {of which the expression is
\begin{equation}\label{adddamp}
\begin{aligned}
\delta\gamma_{b}={\rm{Im}}\left\{ i|G_{\rm{mb}}|^{2}\left[ \chi_{\rm{ma}}(\omega)- \chi_{\rm{ma}}^{*}(-\omega)\right] \right\},
\end{aligned}
\end{equation}
where $G_{\rm{mb}}$ is the effective magnomechanical coupling strength and $\chi_{\rm{ma}}(\omega)=\left[ \chi_{m}^{-1}(\omega)+g_{\rm{ma}}^2 \chi_{a}(\omega) \right]^{-1}$, with $\chi_{m}(\omega)$ ($\chi_{a}(\omega)$) being the natural susceptibility of the magnon (cavity) mode.} The above mechanism works optimally in the resolved sideband limit $\kappa_{\pm} \ll \omega_b$~\cite{Wilson-Rae07,Marquardt07,Genes08}, with $\kappa_{\pm}$ being the polariton linewidths, which is typically well satisfied in the CMM experiments using YIG~\cite{Zhang2016,Potts21,Shen22,Shen23}, benefiting from the low magnonic dissipation rate of the YIG.

\begin{figure}[t]
	\includegraphics[width=450pt]{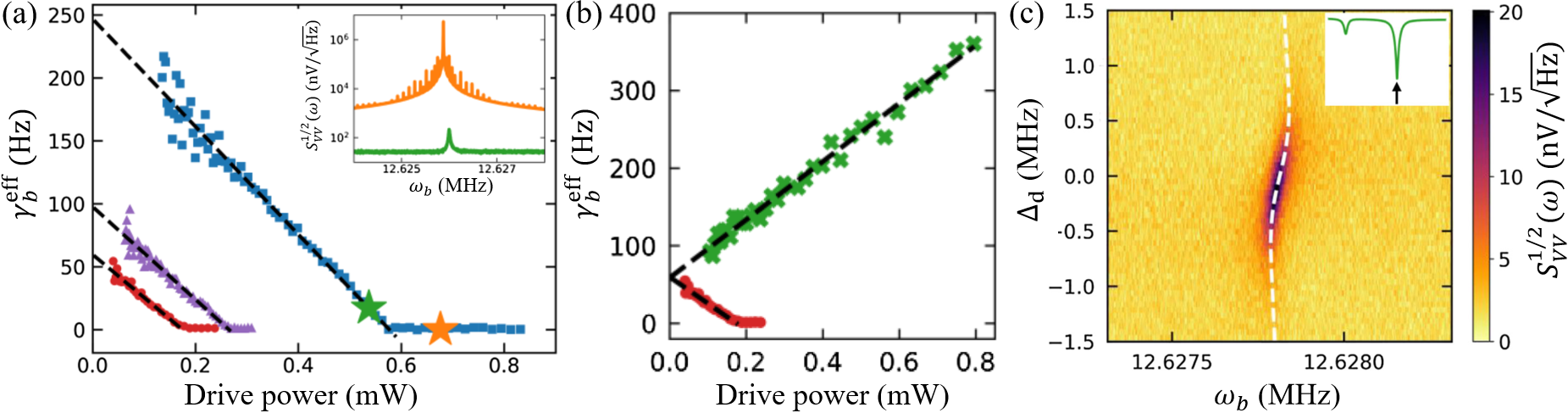}
	\centering
	\caption{(a) [(b)] Effective mechanical damping rate $\gamma_b^{\rm eff}$ versus drive power in the mechanical amplification (cooling) process, when the drive is on resonance with the higher (lower)-frequency polariton. Inset: Mechanical power spectrum for the green marker at drive power of 0.54 mW (the orange marker at power of 0.68 mW). (c) Mechanical power spectrum versus mechanical frequency $\omega_b$ and drive-polariton detuning $\Delta_d$.  The white dashed line is the fitting curve. Inset shows the drive tone. The figures are adapted from Ref.~\cite{Potts21}. 
}
\label{backact}
\end{figure}

Figure~\ref{backact}(a) shows the reduced effective damping rates $\gamma_b^{\rm eff}$ (mechanical linewidths) versus the drive power for three environmental conditions of the YIG sphere placed within a glass capillary, i.e., in air (blue squares), a low pressure ($\sim$15 Torr) of pure helium gas (purple triangles), and vacuum (red circles), corresponding to the natural mechanical damping rate $\gamma_b$ of 247 Hz, 98 Hz, and 59 Hz, respectively. Figure~\ref{backact}(a) is obtained by applying a drive on resonance with the higher-frequency polariton, such that the Stokes scattering is resonantly enhanced. As the drive power increases, a threshold is reached at which the effective damping rate $\gamma_b^{\rm eff}$ approaches zero. By further increasing the power, the mechanical oscillations grow exponentially in time and are ultimately limited by higher-order nonlinear effects. This parametric instability is referred to as phonon lasing.  The inset of Figure~\ref{backact}(a) shows the mechanical power spectrum above (orange) and below (green) the threshold power. When the drive is on resonance with the lower-frequency polariton, the anti-Stokes scattering is resonantly enhanced, leading to a broadened mechanical linewidth (or an increased $\gamma_b^{\rm eff}$) (green crosses in Figure~\ref{backact}(b)). As a comparison, the red circles in Figure~\ref{backact}(b) correspond to the mechanical amplification process as in Figure~\ref{backact}(a). Due to the small magnomechanical cooperativity (slightly greater than 1) achieved in the experiment, the cooling effect is weak and the effective mean phonon number $\bar{n}_b^{\rm eff}$ is only reduced by a few times from the initial large mean thermal phonon number  $\bar{n}_b$.

The mechanical motion can be cooled into its quantum ground state when the cooperativity is significantly improved and the system is at a low bath temperature~\cite{Ding20, Asjad23}. This is a prerequisite for preparing various quantum states in the CMM system~\cite{Li18,Yu20,Li19b,Li19}.  Further, it is shown that magnon squeezing can significantly suppress the magnomechanical Stokes scattering and thus becomes particularly useful in realizing ground-state cooling in the unresolved-sideband regime~\cite{Asjad23}, where the sideband cooling mechanism becomes inefficient. This is the case for the CMM system using other ferromagnetic materials, e.g., CoFeB~\cite{Kansanen21}, which exhibits much stronger magnetostriction but also much larger magnon dissipation. As for mechanical amplification, phonon laser has also been theoretically studied~\cite{Ding19,Xiao21,VG21,Zhang23,Wang23b}, e.g., nonreciprocal~\cite{Xiao21} and exceptional-point-engineered~\cite{Wang23b} phonon laser are suggested.

Another magnomechanical backaction effect is called the ``magnonic spring" effect, in analogy to the ``optical spring" in optomechanics{~\cite{Corbitt07}}. It refers to the phenomenon that the magnomechanical interaction induces a frequency shift of the mechanical mode. {The mechanical frequency shift is given by}
\begin{equation}
\begin{aligned}
{\delta\omega_{b}= -{\rm{Re}}\left\{ i|G_{\rm{mb}}|^{2}\left[ \chi_{\rm{ma}}(\omega)- \chi_{\rm{ma}}^{*}(-\omega)\right] \right\}.}
\end{aligned}
\end{equation}
The ``magnonic spring" effect is revealed in Figure~\ref{backact}(c) for a mechanical resonance of $\omega_b/2\pi =12.6278$ MHz in the partial pressure of helium. It shows that the mechanical oscillator is spring-softened ($\delta \omega_b <0$) for a red-detuned drive and spring-hardened ($\delta \omega_b >0$) for a blue-detuned drive. Though weak, this is the first observation of the ``magnonic spring" effect~\cite{Potts21}. 

Dynamical backaction can, however, be detrimental for some specific applications. A magnomechanical backaction evading experiment has been demonstrated~\cite{Potts23}. A microwave drive tone, tuned between the two polaritons, simultaneously activates both the Stokes and anti-Stokes scatterings. By carefully balancing the scattering rates of the two processes, dynamical backaction effects are eliminated, confirmed by the measured drive-power-independent mechanical linewidth~\cite{Potts23}.

\subsection{Magnomechanical cross-Kerr nonlinearity and mechanical bistability}

As introduced above, the ``magnonic spring" effect causes a frequency shift of the mechanical mode. This frequency shift is typically small, at the level of $0.1$ kHz as observed in the experiment~\cite{Potts21} under a relatively low drive power. However, when the drive power becomes strong, e.g., higher than 10 dBm, a much larger mechanical frequency shift is observed than that is predicted by the dispersive magnomechanical coupling~\cite{Shen22}. This implies that some higher-order nonlinear coupling between magnons and phonons is activated by the strong drive, resulting in a considerable mechanical frequency shift. 

This nonlinear coupling is termed as the magnon-phonon cross-Kerr nonlinearity~\cite{Shen22}, which originates from the magnetoelastic coupling by including the second-order terms in the nonlinear strain tensor~\cite{Landau,Kansanen21}
\begin{equation}\label{f-3}
\epsilon_{ij}=\frac{1}{2}\left ( \frac{\partial u_i}{\partial l_j} +\frac{\partial u_j}{\partial l_i} +\sum_{k}\frac{\partial u_k}{\partial l_i}\frac{\partial u_k}{\partial l_j} \right ),
\end{equation}
where $u_i$ are the components of the displacement vector, {and $l_i = i$ ($i=x,y,z$)}. The first-order terms lead to the conventional dispersive (radiation-pressure-like) interaction Hamiltonian, $\hbar g_{\rm mb} m^\dagger m  \left (b+ b^\dagger\right)/\sqrt{2}$ (Section~\ref{magnomechanicaltheory}). For a not strong drive field, the second-order terms are negligible~\cite{Zhang2016,Potts21}, but they can no longer be neglected when the drive becomes sufficiently strong, yielding considerable magnon-phonon cross-Kerr nonlinearity with the interaction Hamiltonian $\hbar K_{\rm{cross}}m^{\dagger} m  b^{\dagger} b$~\cite{Shen22}, where $K_{\rm{cross}}$ is the cross-Kerr coefficient.

The dispersive magnomechanical coupling gives rise to an effective susceptibility of the mechanical mode~\cite{Shen22}
\begin{eqnarray}
\begin{aligned}
\chi_{\rm{b,eff}}(\omega)
=\left\{ \chi_{b}^{-1}(\omega)-i|G_{\rm{mb}}|^{2}\left[ \chi_{\rm{ma}}(\omega)- \chi_{\rm{ma}}^{*}(-\omega) \right] \right\}^{-1},
\end{aligned}
\end{eqnarray}
where $\chi_{b}(\omega)$ is the natural susceptibility of the mechanical mode, but depends on the modified mechanical frequency $\tilde{\omega}_{b}=\omega_{b}+ K_{\rm{cross}}|M|^{2}$ ($|M|^{2}$ is the magnon excitation number), which includes the cross-Kerr-induced frequency shift. The effective coupling $G_{\rm{mb}}=g_{\rm{mb}}M$, and $\chi_{\rm{ma}}(\omega)=\left[ \chi_{m}^{-1}(\omega)+g_{\rm{ma}}^2 \chi_{a}(\omega) \right]^{-1}$ {as defined in equation~\eqref{adddamp}, but here} the magnon frequency in $\chi_{m}(\omega)$ is modified to be $\tilde{\omega}_{m}=\omega_{m}+2 K_{m}|M|^{2}$, which includes the frequency shift dominantly contributed from the magnon self-Kerr nonlinearity~\cite{YPWang16,YPWang18} caused also by the strong drive, and $K_{m}$ is the magnon self-Kerr coefficient.

\begin{figure}[t]
	\includegraphics[width=400pt]{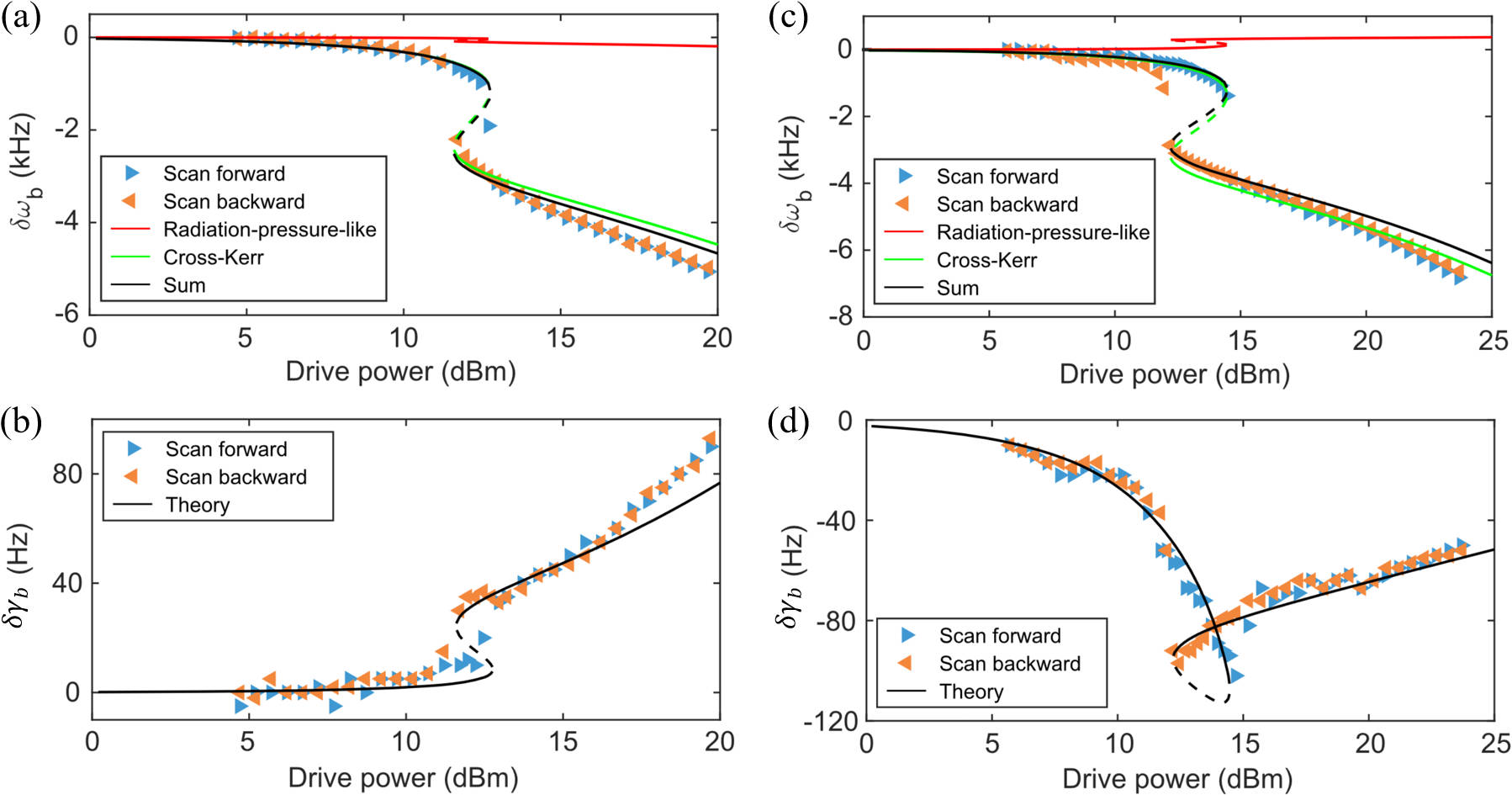}
	\centering
	\caption{Bistable mechanical frequency and linewidth under a red (blue)-detuned drive (a)-(b) [(c)-(d)]. (a) and (c): The mechanical frequency shift versus the drive power. (b) and (d): The mechanical linewidth variation versus the drive power.  The curves are the fitting results. The blue (orange) triangles are the experimental data obtained via forward (backward) sweep of the drive power. The figures are adapted from Ref.~\cite{Shen22}.  }
\label{bist}
\end{figure}

The effective mechanical susceptibility yields a mechanical frequency shift (i.e., the ``magnonic spring" effect)
\begin{equation}\label{freqshift}
\begin{aligned}
\delta\omega_{b}= -{\rm{Re}}\left\{ i|G_{\rm{mb}}|^{2}\left[ \chi_{\rm{ma}}(\omega)- \chi_{\rm{ma}}^{*}(-\omega)\right] \right\}+ K_{\rm{cross}} |M|^{2},
\end{aligned}
\end{equation}
where we write together the frequency shift induced by the cross-Kerr nonlinearity (last term).  In addition, it causes a mechanical linewidth change
\begin{equation}\label{lineshift}
\begin{aligned}
\delta\gamma_{b}={\rm{Im}}\left\{ i|G_{\rm{mb}}|^{2}\left[ \chi_{\rm{ma}}(\omega)- \chi_{\rm{ma}}^{*}(-\omega)\right] \right\}.
\end{aligned}
\end{equation}
{Note that the above equation is different from equation~\eqref{adddamp}, because $\chi_m(\omega)$ is redefined with the modified magnon frequency $\tilde{\omega}_m$.}  
Clearly, the linewidth change is only caused by the dispersive coupling, which can be positive (cooling) or negative (amplification) for a red- or blue-detuned drive, as discussed in Section~\ref{backac}. This differs from the frequency shift where both the dispersive and cross-Kerr couplings contribute, and the latter becomes dominant when the drive is strong. This is clearly seen in Figure~\ref{bist}(a) (Figure~\ref{bist}(c)) for a red (blue)-detuned drive.  As the power grows, the cross-Kerr-induced frequency shift (green curve) increases rapidly, whereas the dispersive-coupling-induced frequency shift (red curve) increases slowly and is generally very small, in the order of $0.1$ kHz. In both the driving situations, the cross-Kerr gives rise to a negative frequency shift because of a negative cross-Kerr coefficient $K_{\rm{cross}}/2\pi=-5.4$ pHz in the experiment~\cite{Shen22}.  By contrast, the frequency shifts by the spring effect are opposite under the red- and blue-detuned drives, in agreement with the finding of Ref.~\cite{Potts21}.

The mechanical linewidth change $\delta\gamma_{b}$ versus the drive power for a red (blue)-detuned drive is shown in Figure~\ref{bist}(b) (Figure~\ref{bist}(d)). As expected, a red (blue)-detuned drive results in a dominant anti-Stokes (Stokes) scattering and thus an increased (a reduced) mechanical linewidth. Note that the triple-resonance condition is not fulfilled in the experiment~\cite{Shen22}. This is because due to the magnon self-Kerr effect, the magnon frequency shifts and thus the splitting of the two polaritons varies by changing the drive power. Therefore, the triple-resonance condition is unlikely to be met as the drive power changes (for fixed cavity frequency $\omega_a$ and coupling $g_{\rm ma}$), but can be achieved for a fixed power.

In Figure~\ref{bist}, both the mechanical frequency and linewidth show a bistable feature with the forward and backward sweeps of the drive power for the two driving situations. This is because both of them originate from the bistable magnon excitation number $|M|^{2}$ (by varying the drive power) induced by the magnon self-Kerr effect~\cite{Shen22,YPWang18}. For both the dispersive-coupling-induced frequency shift and linewidth change, the bistability of $|M|^2$ is mapped to the magnon frequency $\tilde{\omega}_{m}$, then to the polariton susceptibility $\chi_{\rm{ma}}(\omega)$, and eventually to the mechanical frequency and linewidth (cf. equations~\eqref{freqshift} and ~\eqref{lineshift}).

\subsection{Strong-coupling CMM: polariton-mechanics normal-mode splitting}
\label{CPA-SC}

Due to the large frequency mismatch between the magnon and mechanical modes, their coupling is typically weak, leaving the CMM experiments~\cite{Zhang2016,Potts21,Shen22} in the weak-coupling regime. This considerably limits diverse applications of the CMM system, such as quantum states preparation~\cite{Li18,Yu20,Li19b,Li19}. Here, the weak coupling can refer to the following two situations, depending on whether the cavity and magnon modes are strongly coupled: $i$) the effective magnomechanical coupling $G_{\rm mb}$ is smaller than the magnon decay rate $\kappa_m$; or $ii$) the (cavity-magnon) polariton-mechanics coupling $G_{\rm +/-,b}$ is smaller than the polariton decay rate $\kappa_{+/-}$. Since the mechanical damping rate is much smaller, the strong coupling is essentially determined by the relation between the coupling $G_{\rm mb}$ ($G_{\rm +/-,b}$) and the magnon (polariton) decay rate $\kappa_m$ ($\kappa_{+/-}$).

The natural way of achieving the strong coupling is to improve the effective coupling strength $G_{\rm mb}$ or $G_{\rm +/-,b}$, e.g., using a strong drive field.  However, there are restrictions on such improvement: a strong drive field may cause instability~\cite{Anderson55} and unwanted nonlinear effects, such as magnon self-Kerr~\cite{YPWang16,YPWang18} and magnon-phonon cross-Kerr~\cite{Shen22} nonlinearities. One may consider using other ferromagnetic materials, e.g., CoFeB~\cite{Kansanen21}, which exhibit stronger magnetostriction but also significantly increased magnon dissipation. This may end up with a reduced magnomechanical cooperativity and thus may not be an effective means.

\begin{figure}[t]
	\includegraphics[width=450pt]{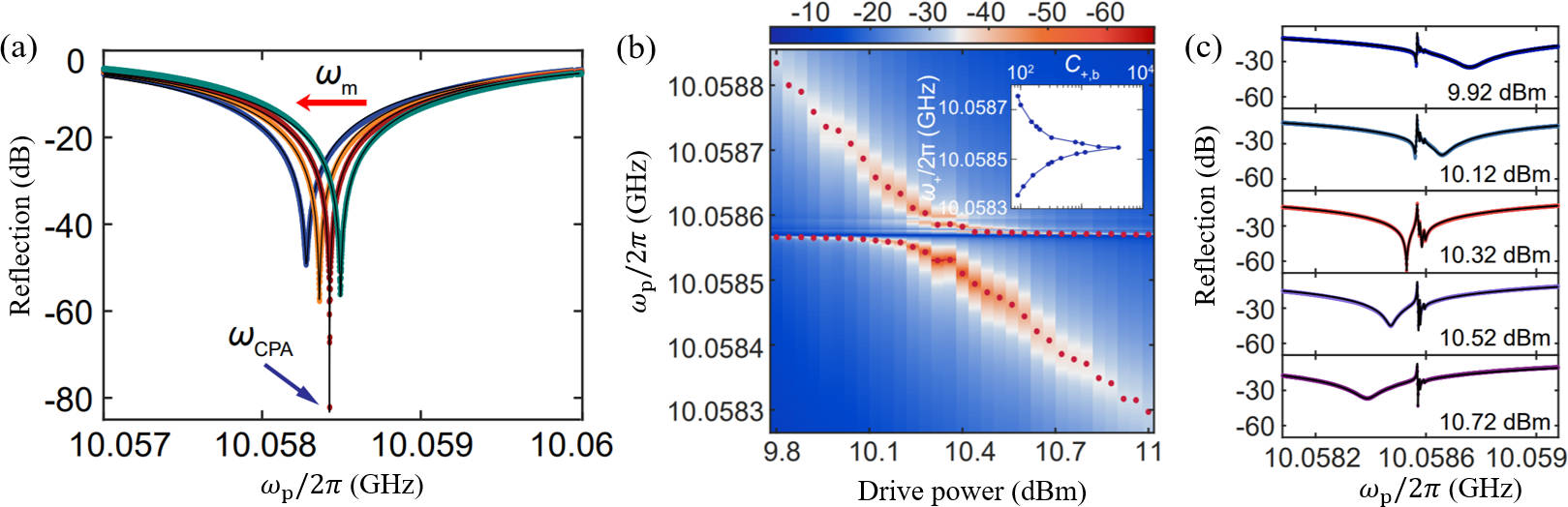}
	\centering
	\caption{(a) Reflection spectra of the higher-frequency polariton. The smallest decay rate of the polariton is achieved at the CPA frequency $\omega_{\rm CPA}$ by tuning the magnon frequency. Polariton-mechanics NMS: (b) Reflection spectra of the polariton-mechanics normal modes versus the drive power.  Red dots correspond to the frequencies of the dips in the reflection spectra. The inset shows the polariton-mechanics cooperativity versus the frequency of the polariton. (c) Cross sections at different drive powers. The figures are adapted from Ref.~\cite{Shen23}.  }
\label{strong}
\end{figure}

An effective approach is offered~\cite{Shen23} by significantly reducing the dissipation rate of the polariton mode exploiting coherent perfect absorption (CPA)~\cite{Chong2010,Cao2011,Cao2012,Chong2017,Zhang17,Yang2021}. Specifically, the cavity is fed by a microwave field to realize an effective gain of the cavity mode. By operating the system at the (cavity) gain-(magnon) loss balance determined by the CPA conditions, i.e., $\kappa_{a}' \equiv \kappa_{e} - \kappa_{\rm int} = \kappa_{m}$ for the resonant case $\omega_{m}=\omega_{a}$~\cite{Shen23}, with $\kappa_{e}$ ($\kappa_{\rm int}$) being the external (internal) decay rate of the cavity, the decay rate of one of the two polaritons (e.g., the higher-frequency polariton) can be reduced to nearly zero. The experiment~\cite{Shen23} realizes the simplest CPA with a single port~\cite{Chong2010,Cao2011,Yang2021}.  Due to the destructive interference between the input and output fields at the cavity port, the amplitude of the cavity output field turns out to be zero, leading to a vanishing dip ($-82$ dB) in the reflection spectrum at the CPA frequency $\omega_{\rm CPA}=\omega_+$ (Figure~\ref{strong}(a)), which is a typical feature of the CPA.   The smallest decay rate of the higher-frequency polariton $\kappa_{+}/2\pi = 140$ Hz is achieved in the experiment~\cite{Shen23}, which is substantiated by the calculation of the Wigner time delay~\cite{Hougne21,LChen21} based on the reflection spectrum of the probe field.  The polariton decay rate is reduced by four orders of magnitude from the normal value $\sim1$ MHz in the cavity magnonic experiments~\cite{Huebl13,Tabuchi14,Zhang14,Goryachev14,Bai15,Zhang15}.

Because of the significantly reduced polariton decay rate and the intrinsically small mechanical damping rate, the polariton-mechanics strong coupling $G_{\rm +,b}>\kappa_{+},\kappa_{b}$  is achieved. Note that the system is actually in the triple cavity-magnon-phonon strong coupling regime, since the polaritons are formed by strongly-coupled cavity photons and magnons.  This leads to the NMS at the mechanical sideband in the spectrum being resonant with the higher-frequency polariton~\cite{Shen23}, as shown in Figure~\ref{strong}(b)-(c).    In the experiment, a red-detuned drive field is applied at frequency  $\omega_{d}/2\pi =10.04762$ GHz and the scattered (anti-Stokes) mechanical sideband is thus at $\omega_{d} + \omega_{b} = 2\pi \times 10.05857$ GHz, where the mechanical frequency $\omega_{b}/2\pi=10.9485~\rm{MHz}$.  By varying the drive power and exploiting the magnon self-Kerr effect, the frequency of the polariton can be continuously tuned (Figure~\ref{strong}(c)).  When the polariton frequency is approaching and then passing through the mechanical sideband, the normal-mode spectrum shows an anti-crossing feature due to the polariton-mechanics strong coupling.  The polariton-mechanics cooperativity, $C_{\rm +,b} = G_{\rm +,b}^2/\left(\kappa_+ \kappa_{b}\right)$, can be up to $C_{\rm +,b} \approx 4\times10^3$ when the polariton nearly resonates with the mechanical sideband (inset of Figure~\ref{strong}(b)), which is improved by three orders of magnitude than the previous CMM experiments~\cite{Zhang2016,Potts21,Shen22}.  Similar NMS at the mechanical sideband has been demonstrated in optomechanical systems, as a sign of the optomechanical strong coupling~\cite{Groblache09,Teufei11,Verhagen12}.

The remarkably improved cooperativity can significantly boost the mechanical cooling efficiency and lower the threshold for achieving phonon lasers~\cite{Potts21}, and makes it possible to prepare magnonic and mechanical quantum states~\cite{Li18,Li19b,Li19}, if further placing the system at a low-temperature environment giving rise to the quantum cooperativity $C_{\rm +,b}^{Q} \equiv C_{\rm +,b}/\bar{n}_b$ much greater than 1.  The experiment~\cite{Shen23} paves the way towards coherent control and measurement of the quantum states of phonons, photons and magnons.

\subsection{High-order sidebands and frequency combs}
\label{High-order-sidebands}

\begin{figure}[t]
	\includegraphics[width=400pt]{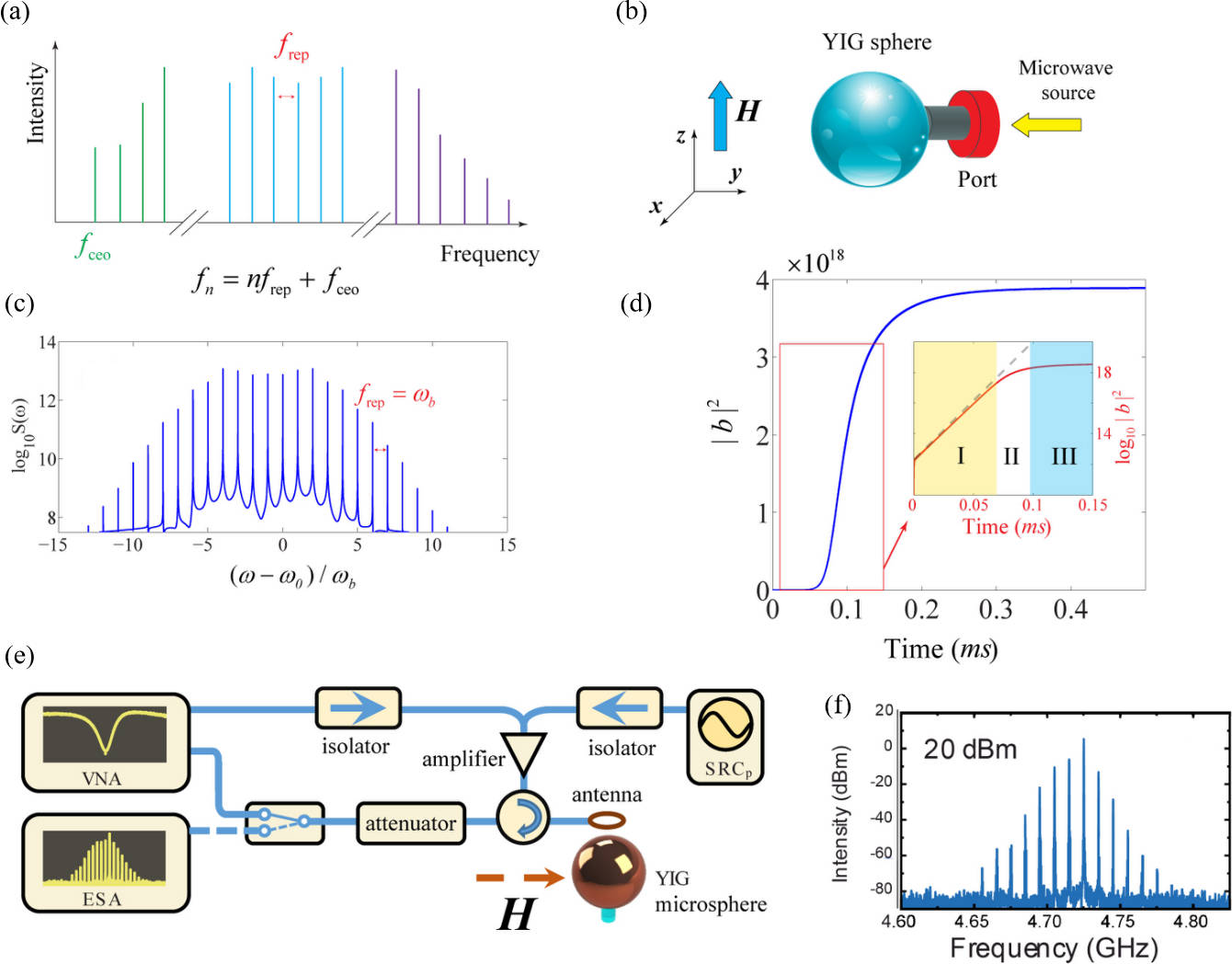}
	\centering
	\caption{(a) Schematic of an OFC with a set of discrete modes that are equally spaced in frequency. (b) Schematic of the generation of MFCs based on the magnetostrictive effect. (c) The magnonic spectrum exhibits a typical frequency-comb structure with the repetition rate $f_{\mathrm{rep}}=\omega_b$ under the drive amplitude over the threshold. (d) Time evolution of the phonon number under a strong drive over the threshold. Inset: phonon number in the logarithmic form versus time. (e) Schematic of the experimental setup for the generation of MFCs via giant mechanical oscillation. (f) The magnonic spectra under the pump power of $20\:\mathrm{dBm}$.
 Figures (a)-(d) are adapted from Ref.~\cite{Xiong23} and Figures (e)-(f) are adapted from Ref.~\cite{Dong23}. }
	\label{fighsg}
\end{figure}

The MMIT and MMIA can be well described by the linearized magnomechanical interaction. Taking into account the nonlinearity of the magnetostriction, it can be used to generate magnonic high-order sidebands (similar to optical high-order sidebands created in optomechanical systems \cite{Xiong18}), with frequencies $\omega_{d}+n \omega_b$, where $\omega_{d}$ is the driving frequency and $n$ is an integer denoting the order of the magnonic sideband.  The efficiency of the magnonic second-order sideband generation \cite{Yang23} can be calculated analytically within the perturbative regime by considering that the generation of the second-order sideband is seeded by the drive field at the first-order sideband. Magnonic high-order sidebands find many applications, e.g., in magnonic information processing and magnonic frequency combs (MFCs).

The concept of the MFC can be regarded as a magnonic counterpart to the optical frequency comb (OFC), which contains a set of discrete modes that are equally spaced in frequency (Figure \ref{fighsg}(a)). As is known, OFCs are vital for and have revolutionized the fields of high-precision frequency metrology and spectroscopy. The extension of frequency comb techniques \cite{Greentree,Xiong23L} to novel physical carriers, such as phonons \cite{Cao14} and magnons \cite{Yan21,Hula22}, is likely to open up new vistas for applications, especially in spectroscopy and sensing. Nevertheless, the efficiency of magnonic sideband generation is very weak due to the intrinsically weak magnomechanical interaction. Hence, strategies are highly needed to enhance the generation of magnonic high-order sidebands.

The sideband generation can be enhanced via phonon lasing (or giant mechanical oscillation). Along this line, the generation of MFCs based on resonantly enhanced magnetostrictive interaction has been proposed \cite{Xiong23}. In Ref.~\cite{Xiong23}, a YIG sphere is placed in the vicinity of a port, which can load a microwave drive field via a loop antenna (Figure \ref{fighsg}(b)). A bias magnetic field is applied to establish the coupling between the drive field and the magnon modes. Magnonic and mechanical dynamics can be substantively modified due to the magnetostrictive effect, which results in degenerate and non-degenerate four-wave mixing and frequency conversion of magnons.   A threshold for the amplitude of the drive field is obtained, and when the drive field is above the threshold, it leads to unstable dynamics where the magnon number grows over time, yielding sufficiently strong magnetostrictive nonlinearity to generate wave mixing and high-order sidebands (Figure \ref{fighsg}(c)). This can be understood from the perspective of phonon lasing. The magnetostrictive interaction leads to the modification of the mechanical damping rate. Below the threshold, the {\it effective} mechanical damping rate is positive and the magnonic dynamics is stable. Whereas the drive over the threshold results in a negative effective damping rate and thus the amplification of the mechanical motions manifested as self-induced oscillations (Figure \ref{fighsg}(d)).

Experimental generation of the MFCs based on the magnetostrictive interaction has been recently realized \cite{Dong23}. In the experiment (Figure \ref{fighsg}(e)), a YIG sphere with the diameter of $623.6\:\mathrm{\mu m}$ is placed in a bias magnetic field, and a microwave field with frequency about $4.725\:\mathrm{GHz}$ is applied via an antenna to excite and drive the magnon mode. For a weak microwave pump with the power below the threshold of $17.4\:\mathrm{dBm}$, the magnonic spectrum contains only the frequency component of the pump field, with no comb lines. When the pump power reaches the threshold, an MFC emerges with a tooth spacing of $\omega_{b}=10.08\:\mathrm{MHz}$ (Figure \ref{fighsg}(f)) and the number of the comb lines increases steadily with the pump power.

{The efficiency of sideband generation can also be enhanced by adopting a two-tone microwave drive~\cite{Liu23} or the cavity-magnon dissipative coupling~\cite{Liu24}.} 
Nonreciprocal sideband generation has been proposed in a spinning microwave magnomechanical system, where a YIG sphere is coupled to a spinning resonator \cite{Wang23}. These proposals provide a pathway for the generation of robust MFCs that may be beneficial for on-chip microwave isolation devices and magnon-based precision metrology.

\subsection{Thermometry}

\begin{figure}[t]
\centering\includegraphics[width=11cm]{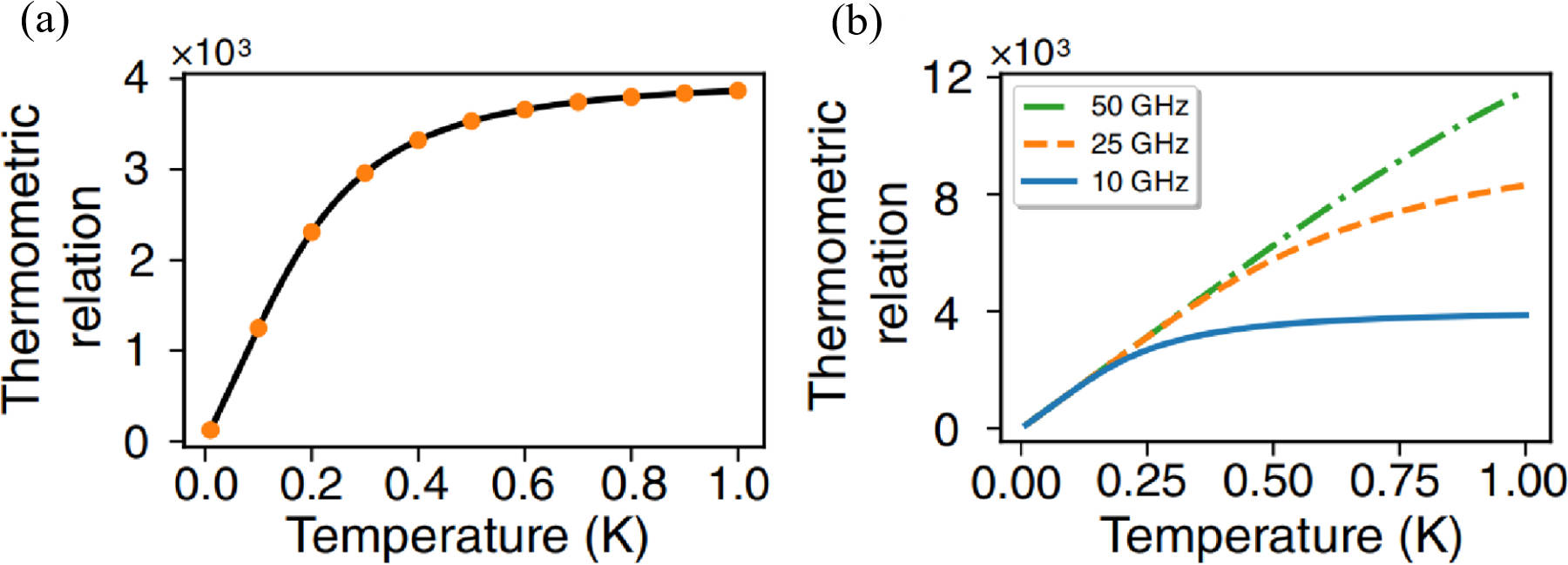} 
\caption{(a) Thermometric relation as a function of bath temperature. The solid line represents the analytical result and the solid circles are numerical simulation. (b) Thermometric relation for different values of the microwave cavity frequency (in resonance with the magnon mode): $\protect\omega_{m}=$10, 25, 50 GHz. The figures are adapted from Ref.~\cite{Potts2020}.}
\label{ther2}
\end{figure}

Many quantum information tasks and quantum technologies require low-temperature environments. In cryogenic environments, accurate thermometry can be
difficult to implement and often requires calibration to an external reference. To solve this problem, Ref.~\cite{Potts2020} proposes a thermometry method using the CMM system at a low bath temperature.  The thermometry is based on the measurement of two correlation spectra of the cavity output field, i.e., the phase-phase autocorrelation and the amplitude-phase correlation spectra.  The temperature of the low-frequency phonon mode (in equilibrium with the bath) is determined by the ratio of the two correlation spectra, i.e., the so-called thermometric relation~\cite{Potts2020}.  The thermometric relation as a function of the bath temperature is shown in Figure~\ref{ther2}(a). It should be noted that the thermometric relation is independent of experimental parameters, e.g., coupling strengths and decay rates. The thermometric measurement is most accurate at low temperatures when the thermal microwave/magnon occupation is less than unity, implying that higher microwave/magnon resonance frequencies are preferred, cf. Figure~\ref{ther2}(b).

Such CMM-based thermometry shows some advantages compared with that employing optomechanical systems. The heating due to optical absorption limits the effective lower temperature range of optical quantum-correlation thermometry to ${\sim}10$ K~\cite{Purdy2017,Meenehan2015}. By contrast, the use of microwave photons, which cause minimal heating of the mechanical system due to their low energy, allows this protocol to be effectively used at temperatures below ${\sim}500$ mK.

\section{Quantum cavity magnomechanics}
\label{quCMM}
 
The nonlinear magnetostrictive interaction provides necessary nonlinearity for preparing various quantum states, e.g., entangled and squeezed states, in the CMM system.  By further exploiting, e.g., external quantum fields, the reservoir engineering technique, and measurements on the microwave field, quantum states of the system can be created or enhanced. In general, the preparation of quantum states requires the system to be at a low bath temperature to reduce the detrimental thermal noises. Otherwise, a cooling mechanism for the mechanical motion should be present.

\subsection{Quantum entanglement}

Many proposals indicate that the CMM system holds the potential to prepare entangled states of magnons and phonons of macroscopic YIG crystals. These  macroscopic entangled states are relevant to the fundamental studies of macroscopic quantum mechanics~\cite{Frowis18}, the boundary between the quantum and classical worlds~\cite{Leggett02,Chen13}, decoherence theories in macroscopic scale~\cite{Bassi13,Weaver18}, etc. Further, the CMM system can also be exploited to prepare entangled states of microwave fields, which find wide applications in quantum information processing~\cite{Lachance-Quirion19,Yuan22}, quantum teleportation~\cite{Furusawa98}, and quantum metrology~\cite{Giovannetti11}. 

\subsubsection{Magnon-photon-phonon entanglement} 
\label{Magnon-photon-phonon entanglement}

The Hamiltonian~\eqref{HHH} leads to the following quantum Langevin equations (QLEs) by including the dissipation and input noise of each mode, which, in the frame rotating at the drive frequency $\omega_0$, are given by
\begin{equation}\label{QLEAA}
\eqalign{
		\dot{a}=&- \left(i \Delta_a + \kappa_a \right) a - i g_{\rm ma} m + \sqrt[]{2\kappa_a}a^{in},  \\
		\dot{m}=&- \left(i \Delta_m + \kappa_m \right) m - i g_{\rm ma} a - i g_{\rm mb} m q + \Omega + \sqrt[]{2\kappa_m}m^{in}, \\
		\dot{q}=&\omega_b p,\,\, \dot{p}=-\omega_b q - \gamma_b p - g_{\rm mb} m^\dagger m + \xi,}
\end{equation}
where $\Delta_a = \omega_a - \omega_0$, and $\Delta_m = \omega_m - \omega_0$.  $a^{in}$ and $m^{in}$ are input noise operators for the cavity and magnon modes, respectively, which are zero-mean and obey the following correlation functions: $\langle j^{in}(t)j^{in\dagger}(t^\prime) \rangle=[N_j(\omega_j)+1]\delta(t-t^\prime)$, $\langle j^{in\dagger}(t)j^{in}(t^\prime) \rangle=N_j(\omega_j)\delta(t-t^\prime)$ $(j = a, m)$. $\xi$ is the Langevin force operator, accounting for the Brownian motion of the mechanical oscillator, which is autocorrelated as $\langle \xi(t)\xi(t^\prime) + \xi(t^\prime)\xi(t) \rangle/2 \simeq \gamma_b [2 N_b(\omega_b)+1] \delta(t-t^\prime)$. Here the Markovian approximation has been taken, which is valid for a large mechanical quality factor $Q_m = \omega_b / \gamma_b \gg 1$. $N_k(\omega_k)=[\exp[(\hbar \omega_k/k_B T)]-1]^{-1}$ $(k = a, m, b)$ are the equilibrium mean thermal photon, magnon, and phonon number, respectively, with $k_B$ the Boltzmann constant and $T$ the bath temperature.

The magnon mode is strongly driven to enhance the dispersive magnomechanical coupling strength, leading to a large amplitude of the magnon mode $| \langle m \rangle| \gg 1$. This allows one to linearize the system dynamics around the large average values. Consequently, the linearized QLEs for the quantum fluctuations can be obtained~\cite{Li18}, which can be expressed compactly in the matrix form of $\dot{u}(t)= A u(t) + n(t)$, where $u(t)=[\delta X_a(t),\delta Y_a(t),\delta X_m(t),\delta Y_m(t),\delta q(t),\delta p(t)]^{\rm T}$, $n(t) = [\sqrt[]{2\kappa_a}X_a^{in},\sqrt[]{2\kappa_a}Y_a^{in},\sqrt[]{2\kappa_m}X_m^{in},\sqrt[]{2\kappa_m}Y_m^{in}, 0, \xi]^{\rm T}$. Here the quantum fluctuations are written in the quadrature form, i.e., $\delta X_O = (\delta O + \delta O^\dagger)/\!\sqrt{2}$, $\delta Y_O = i (\delta O^\dagger - \delta O)/\!\sqrt{2}$ ($O=a,m$), and the noise operators are defined in the same way. The drift matrix $A$ is given by
\begin{equation}\label{AAA}
\eqalign{
		A = \begin{pmatrix}
		-\kappa_a & {\Delta}_a & 0 & g_{\rm ma} & 0 & 0 \\
		-{\Delta}_a & -\kappa_a & -g_{\rm ma} & 0 & 0 & 0\\
		0 & g_{\rm ma} & -\kappa_m & \tilde{\Delta}_m & - G_{\rm mb} &0 \\
		-g_{\rm ma} & 0 & -\tilde{\Delta}_m & -\kappa_m & 0 & 0\\
		0 & 0 & 0 & 0 & 0 & \omega_b \\
		0 & 0 & 0 & G_{\rm mb} & -\omega_b & -\gamma_b\\
	\end{pmatrix},} 
\end{equation}
where $\tilde{\Delta}_m = {\Delta}_m + g_{\rm mb} \langle q \rangle$ is the effective detuning including the frequency shift due to the magnomechanical interaction, and $G_{\rm mb} = i \sqrt{2} g_{\rm mb} \langle m \rangle$ is the effective magnomechanical coupling strength, with $\langle q \rangle = -(g_{\rm mb} / \omega_b) | \langle m \rangle |^2$ and  $\langle m \rangle = \frac{\Omega (i \Delta_a + \kappa_a)}{g_{\rm ma}^2 + (i \tilde{\Delta}_m + \kappa_m)(i {\Delta}_a + \kappa_a)} \simeq \frac{i \Omega \Delta_a}{g_{\rm ma}^2 - \tilde{\Delta}_m {\Delta}_a}$, when $ | \Delta_a |,  | \tilde{\Delta}_m | \gg \kappa_a, \kappa_m$.

Because of the linearized dynamics and the Gaussian nature of the quantum noises, the steady state of the quadrature fluctuations is a three-mode Gaussian state, which can be completely characterized by a $6 \times 6$ covariance matrix (CM) $\cal V$, with its entries defined as ${\cal V}_{ij}=\frac{1}{2} \langle u_i(t)u_j(t^\prime) + u_j(t^\prime)u_i(t) \rangle$ $(i,j=1,2,...,6)$. The steady-state CM $\cal V$ can be obtained straightforwardly by solving the Lyapunov equation~\cite{Li18,Vitali07}. With the CM in hand, bipartite and tripartite entanglement of the system can then be quantified by using the logarithmic negativity $E_N$~\cite{Vidal02,Plenio05} and the minimum residual contangle ${\cal R}_{\tau}^{\rm min}$~\cite{Adesso06,Adesso07}, respectively.

\begin{figure}[t]
	\includegraphics[width=350pt]{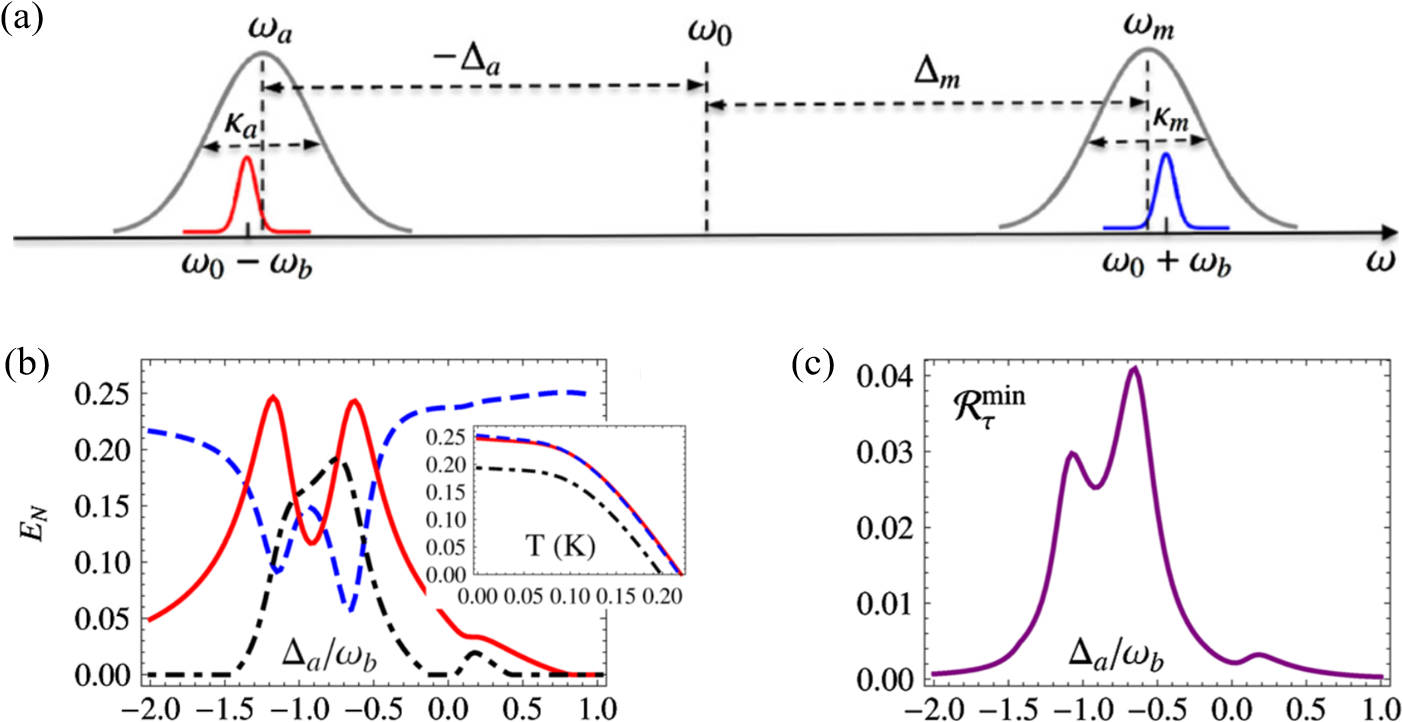}
	\centering
	\caption{(a) Frequencies and linewidths of the system adopted to generate magnon-photon-phonon entanglement. (b) Cavity-magnon (dot-dashed), magnon-phonon (dashed), and cavity-phonon (solid) entanglement versus $\Delta_a$; and versus bath temperature (the inset). $\Delta_a$ is optimized for each bipartite entanglement in the inset. (c) Tripartite entanglement versus $\Delta_a$. The figures are adapted from Ref.~\cite{Li18}. }
	\label{En1}
\end{figure}

A precondition for creating entanglement in the system is the low-frequency mechanical mode being significantly cooled  (close) to its quantum ground state. To this end, a red-tuned microwave field is applied to drive the magnon mode with the detuning $\tilde{\Delta}_m \simeq \omega_b \gg \kappa_m$ (Figure~\ref{En1}(a)). For a relatively weak drive field, this activates the magnomechanical anti-Stokes scattering responsible for cooling the mechanical motion.   When the drive field becomes sufficiently strong,  the magnomechanical coupling  breaks the weak-coupling condition ($G_{\rm mb} \ll \omega_b$) for taking the rotating wave (RW) approximation to obtain the cooling interaction $\propto \delta{m} \delta{b}^\dagger + \delta{m}^\dagger \delta{b}$. Consequently, the counter-RW terms $\propto \delta{m}^\dagger \delta{b}^\dagger + \delta{m} \delta{b}$ in the linearized magnomechanical interaction start to play the role, i.e., the magnomechanical parametric down-conversion (PDC) interaction is activated, yielding the magnomechanical entanglement.     The entanglement  is further distributed to the cavity-magnon and cavity-phonon subsystems when the cavity resonates with the mechanical Stokes sideband, i.e., $\Delta_a \simeq -\omega_b$. In this situation, all bipartite subsystems are entangled, as shown in Figure~\ref{En1}(b).

A more fundamental reason lies in the fact that the two mechanical sidebands, i.e., the Stokes and anti-Stokes sidebands, are entangled via the mediation of the mechanical oscillator, which is involved in both the Stokes and anti-Stokes scattering processes. In the Stokes scattering, the mechanics and the Stokes sideband get entangled due to the PDC, whereas in the anti-Stokes scattering, the mechanics and the anti-Stokes sideband realize an effective state-swap interaction. Therefore, the two sidebands become entangled due to the mechanical mediation.  Since the cavity and magnon modes are, respectively, resonant with the two sidebands, the cavity and magnon modes thus get entangled. Apart from the simultaneous presence of all bipartite entanglements, the system also shares a genuine tripartite entanglement, as witnessed by a nonzero ${\cal R}_{\tau}^{\rm min}$ in Figure~\ref{En1}(c).

The magnon-photon-phonon entanglement can be enhanced by integrating a coherent feedback loop, which contains a controllable beam splitter that determines how much the cavity output field is coherently sent back into the cavity~\cite{Amazioug23}. Since the coherent feedback reduces the cavity decay rate~\cite{Li17}, yielding an increasement of the magnon amplitude and thus an enhanced magnomechanical coupling, the coherent feedback loop improves the tripartite entanglement. 
The entanglement can also be enhanced by utilizing the magnon squeezing~\cite{Ding22} or the microwave squeezing produced by a parametric amplifier~\cite{Hussain22,Zhang22}. 
In addition, controllable transfer of bipartite entanglements is studied in a parity-time-symmetric-like CMM system by introducing an auxiliary active cavity, which provides a gain to the system~\cite{Chen21}. {Very recently, nonreciprocal entanglement~\cite{Chakraborty23, Fan24} and quantum coherence~\cite{Zhang24} have been studied by adopting a spinning resonator~\cite{Chakraborty23, Zhang24} or exploiting the chiral cavity-magnon coupling in a torus-shaped microwave cavity~\cite{Fan24}.}


\subsubsection{Entanglement between two magnon modes}
\label{Entanglement of two magnon modes}

Two magnon modes of two YIG spheres can get nonlocally entangled by exploiting the nonlinear magnetostriction in one YIG sphere and their linear coupling to a common microwave cavity~\cite{Li19}. The system consists of two YIG spheres: one holds a magnon mode and the other supports both a magnon mode and a mechanical vibration mode. This can be realized by adjusting the position of the sphere relative to the direction of the bias magnetic field, since the bare coupling $g_{\rm mb}$ is sensitive to the angle between the direction of the maximum displacement and the bias field \cite{Zhang2016}. The two magnon modes further couple to the same microwave cavity~\cite{Li19}, which plays a key role in distributing the entanglement, as discussed later. The Hamiltonian of the system is the sum of equation~\eqref{HHH} and the following term associated with the second magnon mode ($m$ in equation~\eqref{HHH} is replaced by $m_1$)
\begin{equation}
\eqalign{
		H_{\rm m_2,m_2\text{-}a}/\hbar = \! \omega_{m_2} m_2^\dagger m_2 + g_{\rm m_2 a} \left(a^\dagger m_2 +  a m_2^\dagger \right) ,}
\end{equation}
where $m_2$ and $m_2^\dagger$ ([$m_2$, $m_2^\dagger$] = 1) are the annihilation and creation operators of the second magnon mode, and $g_{m_{2} a}$ is the associated cavity-magnon coupling strength.

\begin{figure}[t]
	\includegraphics[width=450pt]{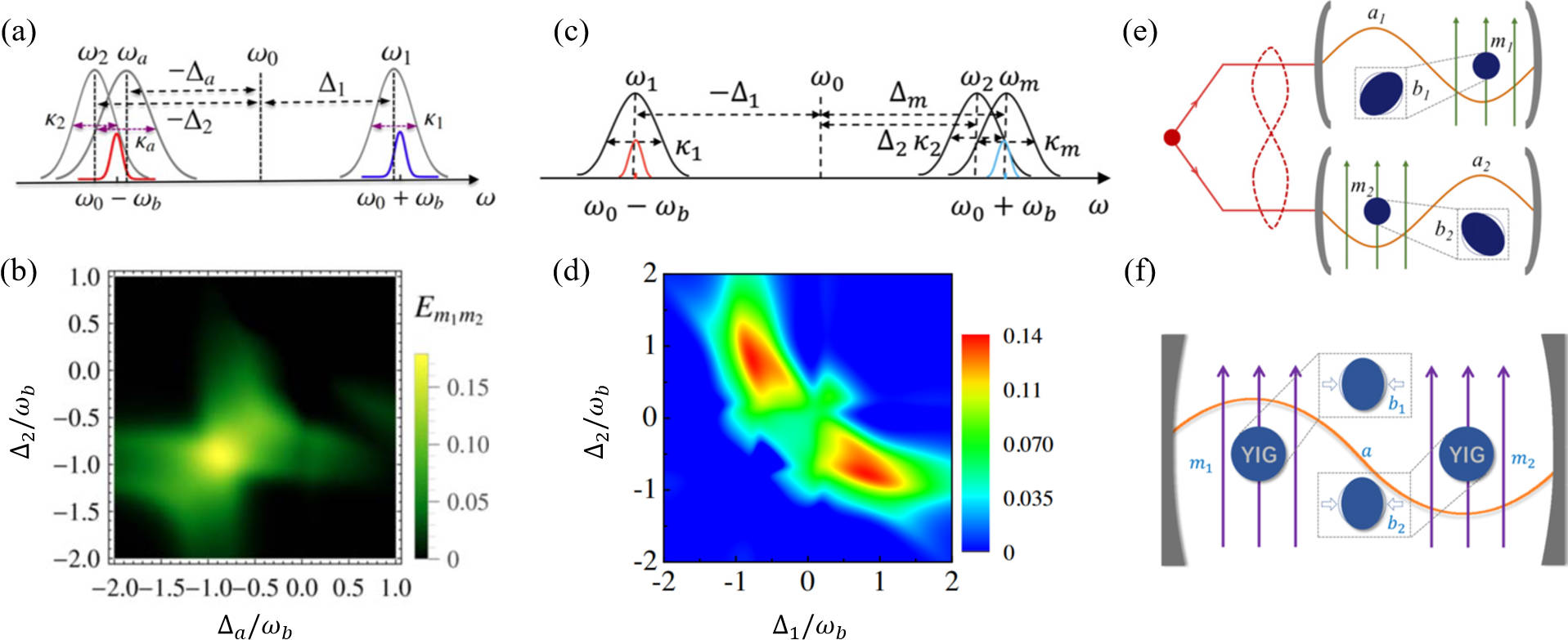}
	\centering
	\caption{(a) Frequencies and linewidths of the system adopted to generate magnon-magnon entanglement. (b) Density plot of the magnon-magnon entanglement versus $\Delta_a$ and $\Delta_2$. (c) Frequencies and linewidths of the system used to entangle two microwave cavity fields. (d) Density plot of the cavity-cavity entanglement vs $\Delta_1$ and $\Delta_2$. (e) Sketch of the dual-CMM system used to entangle two mechanical modes. Two cavities are driven by a two-mode squeezed vacuum field. (f) Sketch of the CMM system used to entangle two mechanical modes of two YIG spheres placed in a common cavity. Figures (a)-(b), (c)-(d), (e) and (f) are adapted from Refs.~\cite{Li19,Yu20,Li21B,Qian23}, respectively. }
	\label{En2}
\end{figure}

The mechanism of entangling two magnon modes can be explained by using the partial result of Ref.~\cite{Li18} for the tripartite CMM system, i.e., the cavity and the magnon mode $m_1$ are entangled under the condition of $\tilde{\Delta}_1 \simeq -\Delta_a \simeq \omega_b$ (here $\Delta_j \equiv \Delta_{m_j}$), cf. Section~\ref{Magnon-photon-phonon entanglement} and Figure~\ref{En2}(a).  Since the second magnon mode $m_2$ also couples to the cavity via the beam-splitter interaction, which realizes a cavity-magnon state-swap interaction, the two magnon modes thus get entangled because of the mediation of the cavity.  The magnon-magnon entanglement is maximized when the cavity and the second magnon mode are resonant, $\Delta_2 \simeq \Delta_a \simeq - \omega_b$ (cf. Figure~\ref{En2}(b)), which is proved to be optimal to realize the cavity-magnon ($m_2$) state transfer~\cite{Li19b}. To sum up, the optimal condition for getting the magnon-magnon entanglement is thus $\tilde{\Delta}_1 \simeq -\Delta_{a,2} \simeq \omega_b \gg \kappa_m$~\cite{Li19}.

\subsubsection{Entanglement between two microwave fields}

Continuous-variable entanglement of microwave fields can be generated by exploiting the nonlinear magnetostrictive interaction and coupling a YIG sphere to two microwave cavities, e.g., of a planar cross-shaped cavity~\cite{Yu20}. The system consists of two microwave cavities and a YIG sphere supporting a magnon mode and a mechanical mode. The Hamiltonian of the system is the sum of equation~\eqref{HHH} and the following term ($a$ is replaced by $a_1$)
\begin{equation}
\eqalign{
		H_{\rm a_2,m\text{-}a_2}/\hbar = \! \omega_{a_2} a_2^\dagger a_2 + g_{\rm m a_2} \big( a_2 + a_2^\dagger \big) \big( m + m^\dagger \big) ,}
\end{equation}
where $a_2$ and $a_2^\dagger$ are the annihilation and creation operators of the second cavity mode, satisfying [$a_2$, $a_2^\dagger$] = 1, and $g_{m a_2}$ denotes the coupling rate between the magnon mode with the second cavity mode.

Figure~\ref{En2}(d) shows that two microwave cavity fields get entangled and the maximum entanglement is achieved when the two cavity fields are, respectively, resonant with the two mechanical sidebands, i.e., $\Delta_1 = - \Delta_2 \simeq \pm \omega_b$, where the ``$\pm$" sign is due to the symmetry of the two cavity fields, and the magnon mode resonant with the blue sideband corresponds to the anti-Stokes process, which significantly cools the phonon mode, thus, removing the main obstacle for observing entanglement. The entanglement originates from the nonlinear magnetostrictive coupling and is distributed to two microwave fields due to the linear magnon-photon coupling. More intuitively, the mechanical motion scatters the microwave driving photons onto two sidebands, which are entangled due to the mechanical mediation as discussed in Section~\ref{Magnon-photon-phonon entanglement}, and if the two cavities are, respectively, resonant with the two sidebands, the two cavity fields get entangled.

\subsubsection{Entanglement between two vibrational modes}

A scheme~\cite{Li21B} is proposed to entangle the mechanical vibration modes of two massive YIG spheres in a dual-CMM system, with each subsystem having the same composition as the model described in Section~\ref{Magnon-photon-phonon entanglement}, shown in Figure~\ref{En2}(e). The two cavities are driven by a two-mode squeezed vacuum microwave field, which entangles the two intracavity fields, and, owing to the cavity-magnon beam-splitter interaction, the two magnon modes thus get entangled. Then, each magnon mode is directly driven by a strong red-detuned microwave field, which activates the magnomechanical state-swap interaction, allowing for the transfer of squeezing from the magnon mode to the phonon mode. Consequently, the two phonon modes of two YIG spheres become entangled.

However, the above scheme needs an external squeezed vacuum field. By fully exploiting the nonlinear magnetostriction, a more energy-saving protocol~\cite{Qian23} is offered without using any external quantum resources. The protocol is based on a hybrid five-mode CMM system, including a microwave cavity mode, two magnon modes, and two mechanical vibration modes, as depicted in Figure~\ref{En2}(f) . The vibration modes and the magnon modes are coupled in a dispersive manner by magnetostriction, and the latter also couple to the microwave cavity via the magnetic-dipole interaction.

 The first step is to entangle the vibration mode $b_1$ and the magnon mode $m_2$. From the result of Ref.~\cite{Li18}, i.e., the mode $b_1$ and mode $a$ are entangled, the mode $b_1$ and mode $m_2$ are thus entangled due to the cavity-magnon ($a$-$m_2$)  state-swap interaction.    Then, switch off the pump on the mode $m_1$ and, simultaneously, turn on a red-detuned pulsed drive on the mode $m_2$. The latter drive activates the magnomechanical beam-splitter interaction, which swaps the magnonic and mechanical states of $m_2$ and $b_2$. Consequently, the previously generated phonon-magnon entanglement is transferred to the two mechanical modes.

\subsection{Quantum steering}

\begin{figure}[t]
\includegraphics[width=450pt]{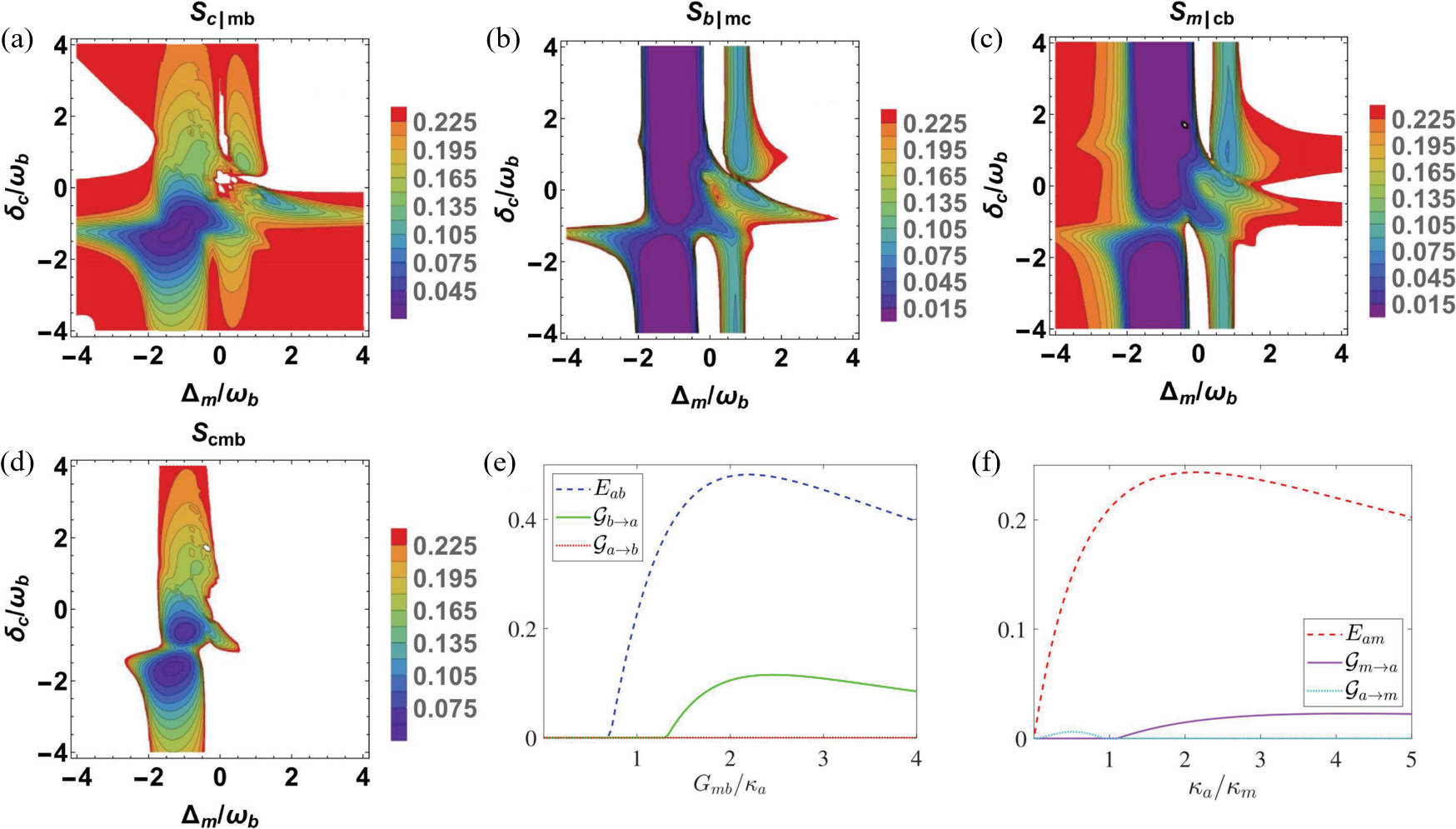}
\centering
\caption{(a)-(d) Bipartite steering $S_{i|jk}$ and genuine tripartite steering $S_{cmb}$ versus magnon- and cavity-drive detunings $\Delta_m$ and $\delta_c$ via weak continuous measurement. Figures are adapted from Ref.~\cite{Tan}. (e) Photon-phonon steering $\mathcal G_{a\leftrightarrows b}$ versus $G_{\rm mb}$; (f) Photon-magnon steering $\mathcal G_{a\leftrightarrows m}$ versus $\kappa_a$. Figures are adapted from Ref.~\cite{Zhang22}.}
\label{f1}
\end{figure}

As another form of quantum correlation that lies in between entanglement and Bell nonlocality \cite{Wiseman}, Einstein-Podolsky-Rosen (EPR) steering characterizes one's capability to control the state of a remote particle via local measurements on his/her particle entangled with the remote one \cite{He22}, i.e., it essentially embodies quantum nonlocality. In Ref.~\cite{Li18}, it is shown that in the CMM system, weak genuine tripartite magnon-photon-phonon entanglement can be achieved in the steady state under a red-detuned (magnon) drive. 
However, for a red-detuned drive, the steady-state bipartite steerings among magnons, photons and phonons are weak, and therefore genuine tripartite steering cannot be achieved. To enhance the bipartite steerings, one solution is to perform the time-continuous homodyne detection on the output field of the driven microwave cavity \cite{Tan}. The measurement considerably enhances the bipartite steerings, such that robust genuine photon-magnon-phonon tripartite steering can be achieved in the steady state. It is shown that the bipartite and genuine tripartite steering in the blue-detuned regime are stronger than those in the red-detuned regime, and they are optimized when the cavity is nearly resonant with the mechanical sideband (Figure \ref{f1}(a)-(d)). This is because, in the absence of the measurement, the stability condition of the system severely restricts the strength of the magnomechanical coupling in the blue-detuned regime. Such a restriction, however, is broken by the measurement, leading to stronger bipartite and genuine tripartite entanglement  \cite{Tan}.

Distinct from Bell nonlocality and entanglement, EPR steering is essentially asymmetric with respect to two observers.  Such an asymmetric and even one-way steering between the cavity mode and the mechanical (magnon) mode in the CMM system can be achieved by the injection of a squeezed (nonclassical) field into the cavity \cite{Zhang22}. The different damping rates of the cavity, mechanical, and magnon modes lead to asymmetric states with respect to the subsystems, thus giving rise to the asymmetric steerings (Figure \ref{f1}(e)-(f)).

\subsection{Squeezed states}

Squeezed states represent a kind of nonclassical states with the noise at certain phases below the vacuum fluctuation. Squeezed states of light or microwave fields find a wide range of applications in continuous-variable quantum information~\cite{Braunstein05sq} and high-precision measurements~\cite{Caves80sq,Giovannetti04sq,Caves81sq,Taylor13sq,Xia23sq}. Squeezed states of magnons and phonons of a large-size ferromagnet are macroscopic quantum states, and the magnon squeezing can also be exploited to improve the detection sensitivity of dark-matter axions~\cite{Flower19sq, Crescini20sq}. Below we briefly introduce a series of proposals for preparing squeezed states of magnons, phonons, and microwave fields by exploiting different mechanisms. 

\subsubsection{Squeezing transfer from external quantum fields}

\begin{figure}[t]
	\includegraphics[width=420pt]{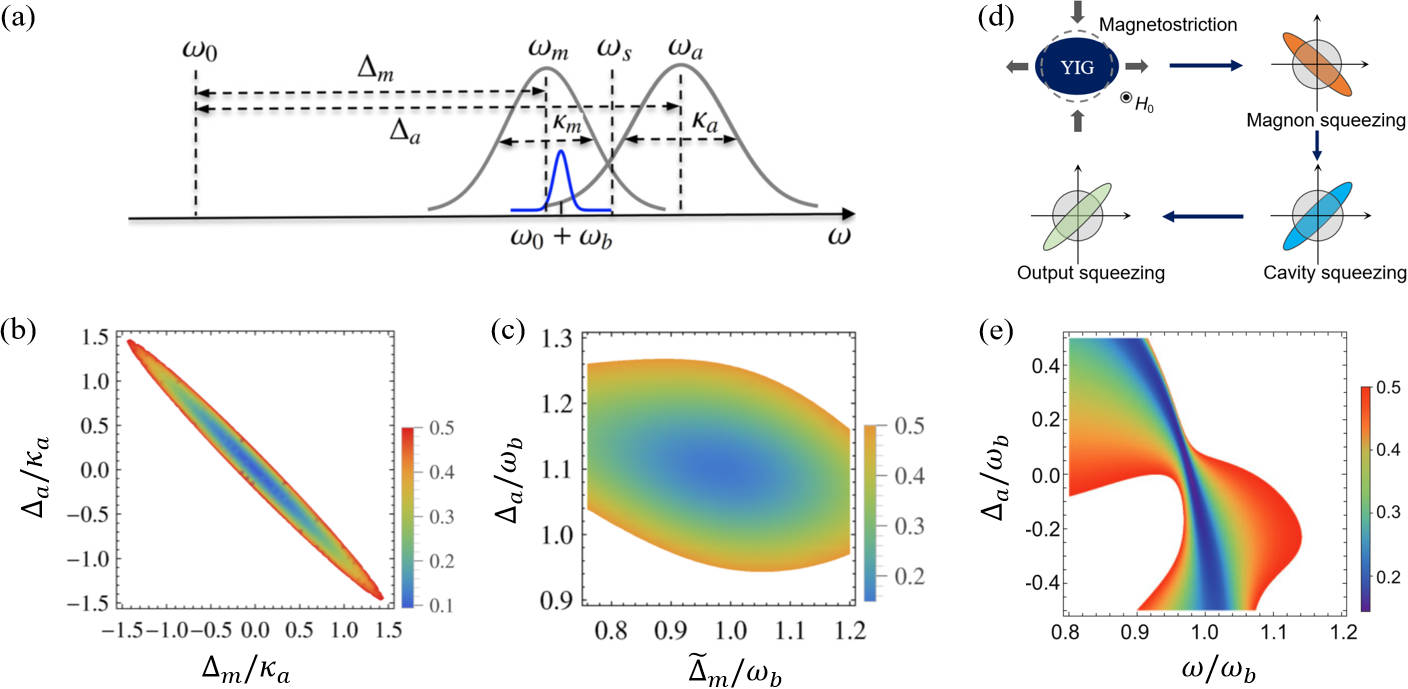}
	\centering
	\caption{(a) Frequencies of the system used to generate squeezed states of magnons and phonons. (b) Variance of the magnon amplitude $\langle \delta x(t)^2 \rangle$ vs $\Delta_m$ and $\Delta_a$. (c) Variance of the mechanical position $\langle \delta q(t)^2 \rangle$ vs $\tilde{\Delta}_m$ and $\Delta_a$. (d) Schematic diagram of the ponderomotive-like magnon squeezing, leading to cavity output field squeezing. (e) Noise spectral density of the output field  $S_W^{\rm out}(\omega)$ versus $\omega$ and $\Delta_a$. The blank areas in (b), (c) and (e) denote the noise above the vacuum fluctuation. Figures (a)-(c) and (d)-(e) are adapted from Refs.~\cite{Li19b,Li23sq}, respectively. }
	\label{Sq1}
\end{figure}

Considering a typical CMM system with the Hamiltonian~\eqref{HHH}, a broadband squeezed vacuum field is injected to the microwave cavity, which shapes the noise properties of the quantum fluctuations of the cavity field, leading to a squeezed cavity field~\cite{Li19b}. The squeezing is then transferred to the magnon mode due to the cavity-magnon beam-splitter (state-swap) interaction. A high transfer efficiency requires the resonant condition $\Delta_a=\Delta_m =0$ (in the frame rotating at the squeezed drive frequency $\omega_s$, cf. Figure~\ref{Sq1}(a)-(b)), the strong coupling, and a much smaller magnon decay rate compared to the cavity decay rate, i.e., $g_{\rm ma} \gg \kappa_a \gg \kappa_m$~\cite{Li19b,Yu20sq}.

Once the magnon mode is squeezed, by further driving it with a strong red-detuned microwave field with frequency $\omega_0$ (Figure~\ref{Sq1}(a)), the magnomechanical state-swap interaction is activated, resulting in the transfer of squeezing from magnons to phonons. The state transfer is optimal when the magnon mode is resonant with the anti-Stokes mechanical sideband, $\Delta_m=\Delta_a =\omega_b$ (redefined with respect to the strong drive frequency $\omega_0$,  cf. Figure~\ref{Sq1}(c)). Note that the red-detuned magnon drive also cools the mechanical mode to its quantum ground state, which is a precondition to achieve the squeezing below vacuum noise~\cite{Li19b}.

\subsubsection{Ponderomotive-like squeezing}

The dispersive (radiation-pressure-like) magnomechanical interaction can be exploited to generate magnon squeezing~\cite{Li23sq}. Specifically, a deformation displacement of the YIG sample is caused by the magnetostriction, proportional to the magnon excitation number, which in turn modulates the phase of the magnon mode, giving rise to a correlation between the amplitude and the phase of the magnon mode. This correlation results in magnon quadrature squeezing. The mechanism is akin to the ponderomotive squeezing of light induced by radiation pressure in optomechanics~\cite{Fabre94sq, Mancini94sq, Brooks12sq, Safavi-Naeini13sq, Purdy13sq, Aggarwal20sq}, so can be termed as the ponderomotive-like squeezing. The squeezing is then transferred to the cavity mode due to the magnon-cavity beam-splitter coupling, and finally to the output field of the cavity, cf. Figure~\ref{Sq1}(d), which can be measured via a homodyne detection.

The optimal condition for squeezing requires a strong red-detuned magnon drive field with a small detuning $0< \Delta_m \,{\equiv}\, \omega_m \,{-}\, \omega_d <\omega_b$, and a remarkable stationary squeezing appears at the mechanical frequency $\omega \simeq \omega_b$ (in the frame rotating at the drive frequency $\omega_d$, Figure~\ref{Sq1}(e)). Such ponderomotive-like squeezing increases with the magnomechanical cooperativity $C_{\rm m,b}=|G_{\rm mb}|^2/(\kappa_m \gamma_b)$, and can be potentially strong~\cite{Li23sq}.

\subsubsection{Reservoir-engineered squeezing}

\begin{figure}[t]
	\includegraphics[width=450pt]{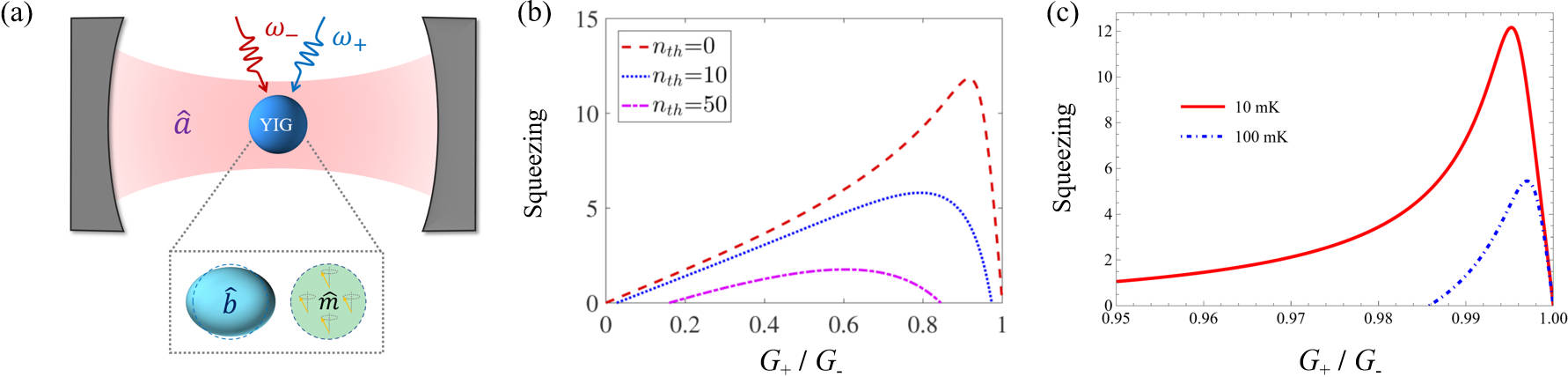}
	\centering
	\caption{(a) Sketch of the reservoir-engineered CMM system used to achieve strong squeezing of mechanical motion and cavity output field. The magnon mode is simultaneously driven by  two microwave fields at frequencies $\omega_\pm$. (b) Mechanical squeezing (in units of dB) versus $G_+/G_-$ at different  mean thermal phonon numbers $n_{\rm th}$. (c) Output field squeezing versus $G_+/G_-$ at different bath temperatures. Figures are adapted from Refs.~\cite{Zhang21sq,Qian23sq}. }
	\label{Sq2}
\end{figure}

The reservoir engineering technique has been successfully applied to the field of optomechanics to prepare squeezed states of mechanical motion~\cite{Tan13,Kronwald13sq, Wollman15sq, Pirkkalainen15sq, Lecocq15sq}.  In view of the similarity between the opto- and magnomechanical interactions, stationary mechanical squeezing can also be obtained by simultaneously driving the magnon mode with both red- and blue-detuned microwave fields~\cite{Zhang21sq,Qian23sq}, cf. Figure~\ref{Sq2}(a).

The two-tone drive can be described by the driving Hamiltonian $H_{\rm{dri}} / \hbar = \left(\Omega_{+}e^{-i\omega_+t}+\Omega_-e^{-i\omega_-t}\right) m^\dagger + \mathrm{H.c.} $, where $\Omega_+$ and $\Omega_-$ are the Rabi frequencies associated with the two drive fields at frequencies $\omega_{\pm}=\omega_m \pm \omega_b$, respectively. The two-tone drive leads to the steady-state mean value of the magnon mode being approximated as $\langle m \rangle \approx \langle m_+ \rangle e^{-i\omega_+t} + \langle m_- \rangle e^{-i\omega_-t}$, where $\langle m_\pm \rangle = \Omega_\pm/\left(\omega_{\pm}-\omega_m + i\kappa_m-\frac{g_{\rm{ma}}^2}{\omega_\pm-\omega_a+i\kappa_a}\right)$. The linearized QLEs of the quantum fluctuations, in the interaction picture with respect to $\hbar \omega_m(a^\dag a + m^\dag m) + \hbar \omega_b b^\dag b$, are given by
\begin{equation}\label{sq3_QLEs}
\eqalign{
\delta\dot{a} & = -\kappa_a \delta a -ig_{\rm{ma}}\delta m +\sqrt{2\kappa_a} a^{in},\\
\delta\dot{b} & = -\gamma_b \delta b - i\left(G_- + G_+ e^{2i\omega_bt} \right) \delta m  -i \left(G_+ + G_- e^{2i\omega_bt}\right) \delta m ^{\dagger} + \sqrt{2\gamma_b} b^{in},\\
\delta\dot{m} & = -\kappa_m \delta m - ig_{\rm{ma}}\delta a - i\left(G_- {+}\,G_+ e^{-2i\omega_bt} \right) \delta b  - i\left(G_+ {+} \,G_- e^{2i\omega_bt} \right) \delta b^{\dagger} \!\! + \!\!\sqrt{2\kappa_m} m^{in},
}
\end{equation}
where the resonant case $\omega_a=\omega_m$ is considered and $G_{\pm}=g_{\rm{mb}}\langle m_{\pm}\rangle$ are the effective magnomechanical couplings associated with the two drive fields. The squeezing or noise properties can be investigated by solving the dynamical covariance matrix of the system~\cite{Zhang21sq}.

The system dynamics can be conveniently solved by taking the RW approximation under the condition of $\kappa_{a(m)}, g_{\rm{ma}}, G_{\pm}\ll\omega_b$, which allows us to neglect the time-dependent coupling terms in equation~\eqref{sq3_QLEs}. This then leads to the effective coupling between the magnon and mechanical modes described by $\delta m^\dagger \left(G_+\delta b^\dagger + G_- \delta b\right) + \rm{H.c.}$, which can be rewritten as ${\cal G} \left( \delta m ^\dagger \delta B \right) + \rm{H.c.}$, with ${\cal G} = \sqrt{G_-^2-G_+^2}$, and $\delta B = \cosh r \,\delta b +\sinh r \, \delta b^\dagger$ being the Bogoliubov mode, where $r =\frac{1}{2} \ln \left(\frac{G_-+G_+}{G_--G_+}\right) $ is the squeezing parameter. It describes a beam-splitter interaction between the magnon mode and the Bogoliubov mode, and can thus be used to cool the Bogoliubov mode. Since the ground state of the Bogoliubov mode corresponds to the squeezed vacuum state of the mechanical mode, mechanical squeezing can be achieved by cooling the Bogoliubov mode. For the stability reason, $G_+<G_-$ is required. It is worth noting that for a fixed $G_-$, by increasing $G_+$ (via increasing the blue-detuned drive field) the squeezing parameter $r$ grows, but this simultaneously reduces the cooling rate ${\cal G}$ of the Bogoliubov mode. Consequently, an optimal ratio of $G_+/G_-$ exists for the mechanical squeezing as a result of the trade-off between the above two effects, cf. Figure~\ref{Sq2}(b).

Despite the above analysis for the mechanical squeezing, the two-tone drive also induces the magnon squeezing and consequently the intracavity field squeezing, but the degrees of squeezing of both the magnon and intracavity field are very small~\cite{Zhang21sq}. Instead, Ref.~\cite{Qian23sq} studies the squeezing of the cavity output field and shows that the squeezing can, however, be significantly large (Figure~\ref{Sq2}(c)). The squeezed output field can be readily accessed and thus can be directly applied to quantum information science and quantum metrology.

\subsection{Quantum memory}

\begin{figure}[t]
\includegraphics[width=450pt]{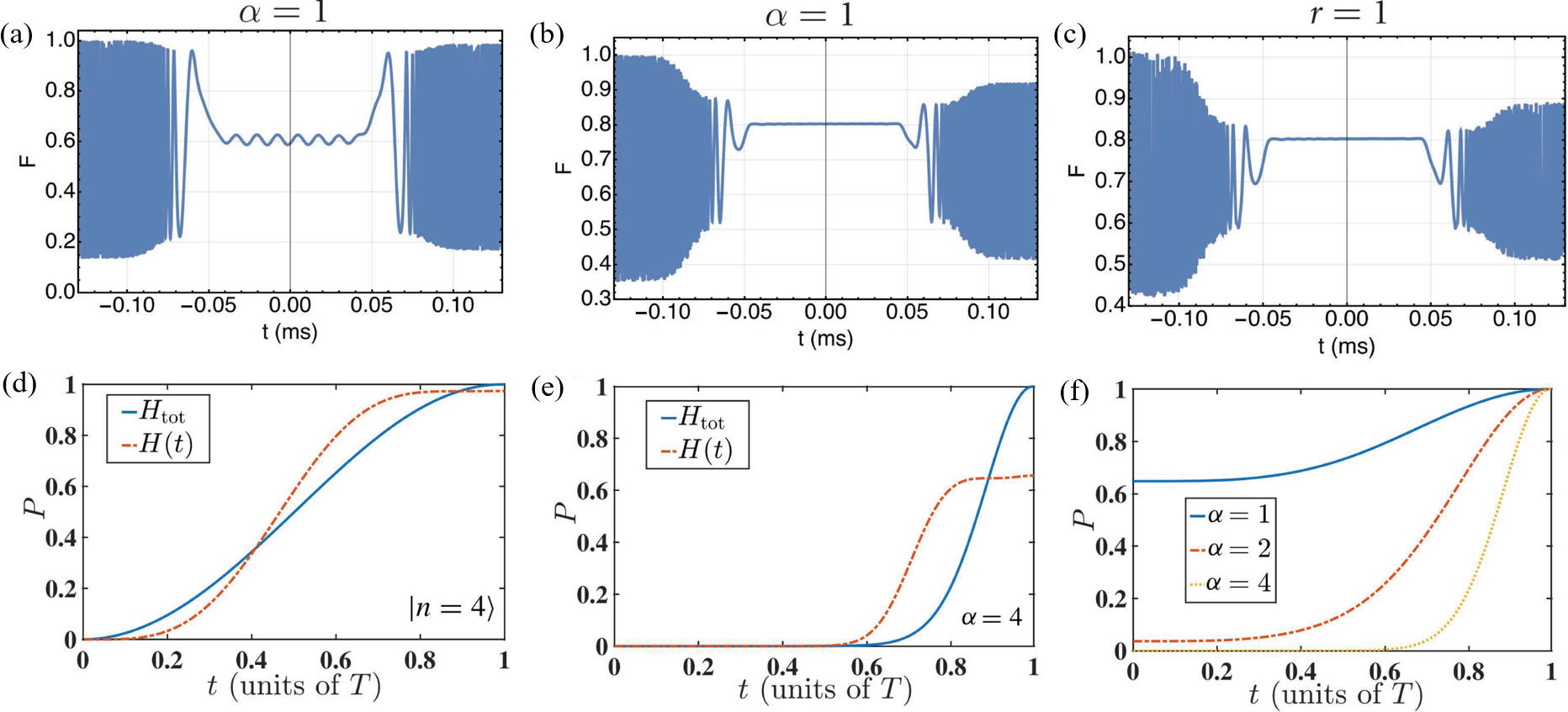}
	\centering
\caption{Transfer fidelities for a cavity input coherent state (a), cat state (b) and squeezed state (c) via modified STIRAP.  Figures are adapted from Ref.~\cite{Sar}. The target-state population of the phonon mode for the initial (d) Fock state $|n=4\rangle$ and (e) cat state of $\alpha=4$ in the TQD protocol, and the initial (f) cat states of  $\alpha=1,2,4$ in the LR-invariant approach. Figures are adapted from Ref.~\cite{Jing2}.}
\label{f3}
\end{figure}

Due to their low damping rates, the mechanical modes of the CMM system can serve as quantum memory for microwave or magnonic quantum states. Further, by exploiting the unique advantages of the CMM system, e.g., tunable magnon frequency and linearized magnomechanical coupling, strong photon-magnon coupling, etc, the CMM system provides a promising platform to efficiently realize the photon-magnon-phonon state transfer. In Ref.~\cite{Jing1}, the magnon-mediated photon-to-phonon transfer in non-Markovian environments is studied, where the cavity and mechanical modes are assumed to be in structured environments with an Ohmic- and $1/f$-like spectra, respectively. It shows that the fidelity of the photon-to-phonon state transfer can be considerably enhanced due to the presence of the non-Markovian environments.

Differently, Ref.~\cite{Sar} adopts the stimulated Raman adiabatic passage (STIRAP) \cite{Berg} to achieve high-fidelity state transfer between the cavity and mechanical modes, where the magnon-phonon coupling and the detunings of the cavity and magnon modes with respect to the drive field are designed to be time-dependent.
Figure \ref{f3}(a)-(c) show the instantaneous fidelities for the state transfer and retrieval processes for an initial coherent state $|\alpha\rangle$, Schr\"{o}dinger's cat state $\mathcal N(|\alpha\rangle+|-\alpha\rangle)$ ($\mathcal N$ as the normalization factor), and squeezed state $e^{(r a^2-r^*a^{\dag 2})/2}|0\rangle$ ($r$ as the squeezing parameter), respectively, from the cavity to the mechanical mode and back to the cavity mode. For all coherent states, the retrieval fidelity is nearly unit. For cat and squeezed states, even though the retrieval fidelity is rather high, it no longer reaches unit and decreases with the increase of $\alpha$ and $r$.

In addition, shortcut-to-adiabatic (STA) protocols are provided in Ref.~\cite{Jing2} to achieve the high-fidelity state transfer. The STA protocols are based on the counterdiabatic Hamiltonian for transitionless quantum driving (TQD) \cite{Ode} and the Levis-Riesenfeld (LR) invariant for inverse engineering~\cite{Lewis}, respectively. It is shown that the TQD approach manifests its power for a larger cat state by achieving a perfect transfer population and the LR-invariant-based approach can further achieve a perfect transfer for any superposed states (Figure \ref{f3}(d)-(f)).

\section{Hybrid systems based on magnomechanics: optomagnomechanics and others}
\label{Optomag}

\subsection{Optomagnomechanics}

\begin{figure}[t]
	\includegraphics[width=1\linewidth]{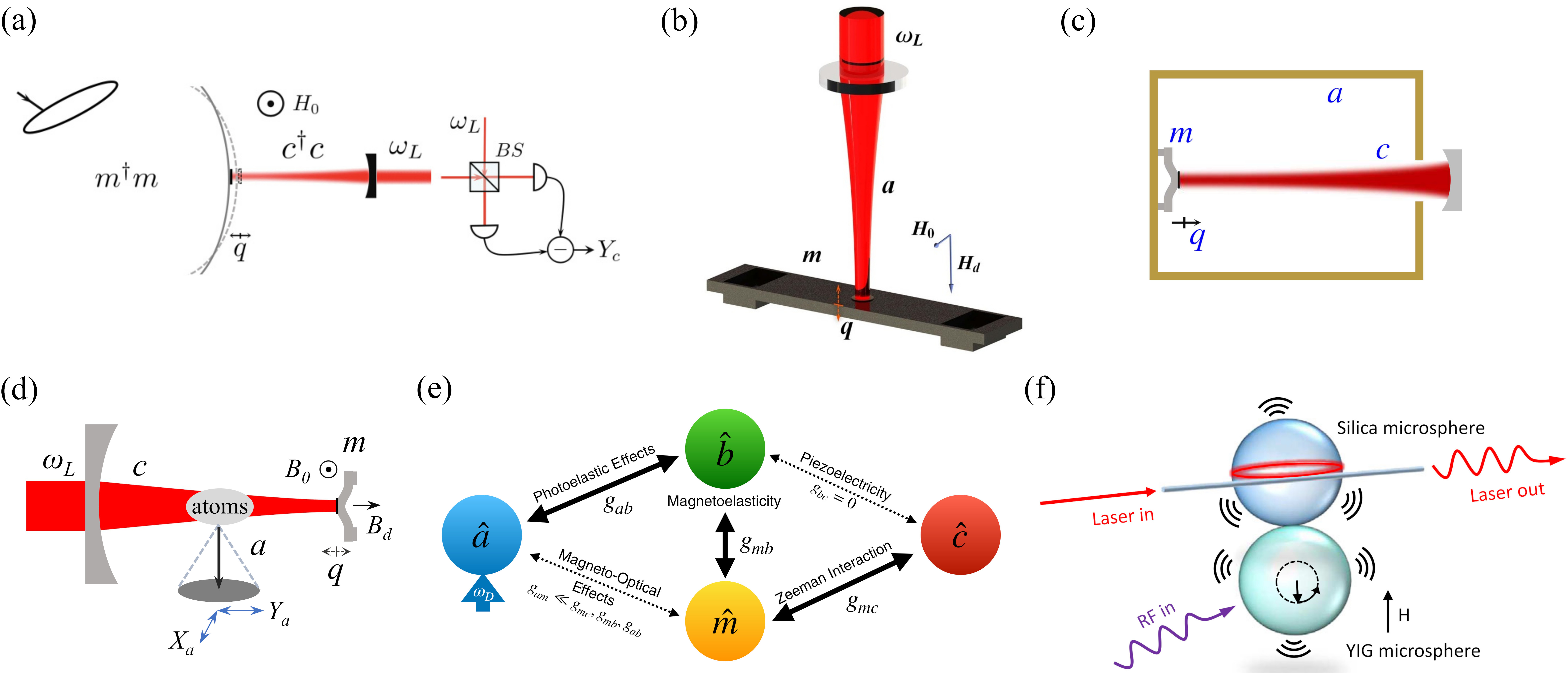}
	\caption{Hybrid systems based on magnomechanics. (a)-(d) Hybrid OMM configurations adopted in Refs. \cite{Fan23,Fan2022,Fan2023B,Fan2023C} to realize magnon population detection~\cite{Fan2022}, magnon state readout and optomagnonic entanglement~\cite{Fan23}, microwave-optics entanglement~\cite{Fan2023B}, and magnon-atomic ensemble entanglement~\cite{Fan2023C}. (e) Microwave-to-optics transduction protocol using a linear coupling \cite{Engelhardt2022}. (f) Hybrid system combining magnomechanics with optomechanics to establish coherent couplings among phonons, magnons and photons \cite{Shen2022}.  Figures (a)-(f) are adapted from Refs.~\cite{Fan2022,Fan23,Fan2023B,Fan2023C,Engelhardt2022,Shen2022}, respectively.}
	\label{FIG1}
\end{figure}

It is quite natural to couple the magnetostriction-induced mechanical displacement with an optical cavity via the radiation pressure~\cite{Aspelmeyer14}. This then forms a hybrid system consisting of magnons, vibration phonons and optical photons, which is termed as {\it optomagnomechanics} (OMM). The OMM system contains both the magno- and optomechanical interactions, which are both a dispersive type (in the case of $\omega_b \ll \omega_m $) and thus provide rich nonlinearities for, e.g., preparing various quantum states in the system~\cite{Fan23,Fan2023B,Fan2023C}. 
A promising physical realization could adopt a micron sized YIG bridge structure \cite{Heyroth2019}, which supports a magnon mode with the frequency in gigahertz and a mechanical vibration mode with the frequency ranging from tens to hundreds of megahertz, thus possessing a dominant dispersive magnomechanical coupling (Section \ref{magnomechanicaltheory}). To increase the $Q$ factor of the optical cavity, a small high-reflectivity mirror pad could be attached to the surface of the magnomechanical resonator~\cite{Fan23} (Figure \ref{FIG1}(a)-(d)).
The Hamiltonian of the OMM system is given by~\cite{Fan23,Fan2022}.
\begin{equation}
	\label{OMMHamitinian}
	\eqalign{
	H/\hbar=&\sum_{j=c,m} \omega_j j^\dagger j+\frac{\omega_b}{2}\left( q^2+p^2\right) -g_{\rm cb} c^\dagger cq+g_{\rm mb} m^\dagger  mq+ H_{\mathrm{dri}}/\hbar,
	}
\end{equation}
where $c$ ($c^\dagger$) is the annihilation (creation) operator of the optical cavity mode with frequency $\omega_c$. {The bare optomechanical coupling strength $g_{\rm cb}$ is in the range of $10^2$-$10^3$ Hz for a micron-size mirror~\cite{Kleckner06, Favero07}. The bare magnomechanical coupling $g_{\rm mb}$ is about tens of Hz for a micron-size YIG bridge~\cite{Fan23} estimated using the theory of Ref.~\cite{Kansanen21}.} The driving Hamiltonian $H_{\mathrm{dri}}=i\hbar \epsilon(c^\dagger e^{-i\omega_L t}-c e^{i\omega_L t})+i\hbar\Omega(m^\dagger e^{-i\omega_0 t}-m e^{i\omega_0 t})$, where $\epsilon$ denotes the coupling between the cavity and the driving laser at frequency $\omega_L$.

It is suggested that the OMM configuration can be used to optically measure the magnon population of the ferromagnet \cite{Fan2022}. Specifically, by resonantly driving the cavity with a weak laser field and under appropriate conditions, a linear relation can be established between the magnon population $N_m=\left|\left\langle m\right\rangle \right|^2$ and the optical phase quadrature $\left \langle Y_c \right \rangle$ via the mediation of the mechanical displacement, i.e., $\left \langle Y_c \right \rangle \simeq -4\sqrt{2}\frac{\epsilon g_{\rm cb} g_{\rm mb}}{\kappa_c^2\omega_b} N_m$\cite{Fan2022}, with $\kappa_c$ being the cavity decay rate. The excellent linear dependence indicates that the magnon population of the ferromagnet can be read out in the phase quadrature of the optical field, e.g., by homodyning the cavity output field (Figure \ref{FIG1}(a)).

The OMM system also enables many quantum protocols, which may find wide applications in quantum information science and technology. For example, by driving the magnon (cavity) mode with a red-detuned microwave (laser) field to activate the magnomechanical (optomechanical) state-swap interaction, the magnon-phonon-photon quantum state transfer can be realized. Consequently, the magnonic state can be read out in the output field of the cavity via the mechanical transduction~\cite{Fan23} (Figure \ref{FIG1}(b)). Instead, if the red-detuned  magnon drive is replaced with a blue-detuned drive to activate the magnomechanical PDC interaction, stationary optomagnonic entanglement can be created~\cite{Fan23}.  Under these conditions, if by further coupling the magnon mode to a microwave cavity via the beam-splitter interaction~\cite{Tabuchi14,Zhang14} (forming an extended microwave cavity-OMM system, Figure \ref{FIG1}(c)), the above generated optomagnonic entanglement \cite{Fan23} can then be distributed to the microwave cavity, yielding stationary microwave-optics entanglement \cite{Fan2023B}. The microwave-optics entanglement finds particularly important applications in building a hybrid quantum network~\cite{Sahu23}. Employing the microwave cavity-OMM system, an entanglement transfer scheme has been suggested to entangle two microwave fields~\cite{Di23}.  Based on the system used in Ref.~\cite{Fan23}, if further coupling the optical cavity to an ensemble of two-level atoms (Figure \ref{FIG1}(d)), macroscopic entanglement between magnons and atoms and genuine tripartite atom-magnon-phonon entanglement can be generated \cite{Fan2023C}.
Recently, an alternative protocol has been provided for entangling magnons and atoms based on the CMM system by directly coupling the microwave cavity to an atomic ensemble~\cite{Wu2023}. 

The above protocols consider the mechanical frequency to be much lower than the magnon frequency. When the size of the magnomechanical system is significantly reduced (considering, e.g., thin YIG films), the mechanical frequency becomes high and close to the magnon frequency~\cite{Godejohann20,An20,Xu21}. In this regime, the magnomechanical interaction is dominated by the (strong) linear coupling (Section \ref{magnomechanicaltheory}). Based on this coupling, a microwave-to-optics conversion protocol is provided~\cite{Engelhardt2022} exploiting the state-swap interaction in the microwave-magnon, magno- and optomechanical interactions (Figure \ref{FIG1}(e)).

\subsection{Combining magnomechanics with optomechanics via mechanics-mechanics coupling}

Another approach to combining magno- and optomechanics is via the direct mechanics-mechanics coupling, implemented by placing the two systems in direct physical contact (Figure \ref{FIG1}(f)). This hybrid system can achieve coherent couplings among magnons, vibration phonons and optical photons~\cite{Shen2022}.  The Hamiltonian of such a system reads 
\begin{align}
	H/\hbar=\sum_{k=c,m,b_1,b_2} \omega_k k^\dagger k +g_{\rm mb} m^\dagger  m \left(b_1+b_1^\dagger \right)+g_{b} \left(b_1^\dagger b_2+b_2^\dagger b_1 \right)+g_{\rm cb} c^\dagger c \left(b_2+b_2^\dagger \right),
\end{align}
where $k\,{=}\,c,m,b_1,b_2$ ($k^\dagger$) are the annihilation (creation) operators of the optical cavity, magnon, and two mechanical modes, respectively, $\omega_k$ are the corresponding mode frequencies, and  $g_b$ denotes the linear coupling between the two mechanical modes of close frequencies.  In such a system, when the magnon (optical cavity) mode is driven by a red-detuned microwave (laser) field, the magnomechanical (optomechanical) beam-splitter interaction is activated, which enables the microwave-to-optics conversion via the mechanical mediation. Moreover, the interference effect between the two mechanical motions is also observed by applying a red- (blue-)detuned drive field to the magnon (cavity) mode~\cite{Shen2022}.

\section{Conclusion and outlook}
\label{conc}

In conclusion, we have reviewed both experimental and theoretical developments in the emerging field of CMM. The experimental observations of the MMIT (MMIA), magnomechanical dynamical {backaction}, magnon-phonon cross-Kerr nonlinearity, and magnetostrictively induced MFCs are the typical characteristics of the CMM system possessing a nonlinear magnetostrictive interaction. The system has also recently reached the strong-coupling regime and the polariton-mechanics NMS has been observed. As for the theoretical investigations, the progress is much ahead of the experimental studies and many proposals for preparing various quantum states of magnons, phonons, and microwave photons have been offered. These theoretical findings provide beneficial guidance for future experimental development.  Despite the above achievements, there is still enormous potential for the development in this field.  The quantum-state protocols indicate that the magnomechanical effective coupling and cooperativity should be significantly improved in order to prepare any quantum states of the system, which is likely the most important direction for future experiments.  A landmark achievement would be the generation of real macroscopic quantum states in the laboratory. Besides, the CMM system also offers an ideal platform for nonlinear researches by exploiting rich nonlinearities from the magnon-phonon coupling and the YIG.  

Below, we provide an outlook for future research directions in this field, including non-Hermitian, $\mathcal{PT}$- and anti-$\mathcal{PT}$-symmetry, and exceptional-point-engineered CMM, as well as self-sustaining dynamics and synchronization phenomena by fully exploiting the nonlinearity of the magnetostriction. At last, we introduce some other magnomechanical platforms beyond the frameworks set by Refs.~\cite{Lachance-Quirion19,Yuan22,Rameshti22}, which, however, are not the focus of this review.

\subsection{Non-Hermitian and $\mathcal{PT}$-symmetry CMM}

Non-Hermitian physics and parity-time ($\mathcal{PT}$) symmetry have attracted considerable attention over the past decade~\cite{El-Ganainy18,Ashida20,Bergholtz21}. Combining non-Hermitian physics with the novel CMM system represents an important research direction and could considerably enrich the research content in this field.


\subsubsection{$\mathcal{PT}$-symmetric CMM}

The $\mathcal{PT}$-symmetric CMM system can be constructed by realizing an {\it effective} gain of the microwave cavity mode with the gain rate $\kappa_a'$, e.g., by exploiting the CPA~\cite{Shen23,Zhang17} (Section~\ref{CPA-SC}).  Due to the intrinsically weak bare magnomechanical coupling $g_{\rm mb}$, for a not very strong drive, the conditions for realizing the $\mathcal{PT}$-symmetric CMM system are almost identical with those for realizing the $\mathcal{PT}$-symmetric cavity-magnon system, i.e., the cavity and the magnon mode being resonant $\omega_a=\omega_m$ and the cavity gain rate and the magnon decay rate being balanced $\kappa_a' = \kappa_m \equiv \kappa$~\cite{Zhang17}.  By diagonalizing the Hamiltonian, one finds that $g_{\rm ma}=\kappa$ corresponds to the $\mathcal{PT}$ phase transition point, i.e., the exceptional point (EP), of the transition from the $\mathcal{PT}$-symmetric phase ($g_{\rm ma}>\kappa$) to the broken-$\mathcal{PT}$-symmetric phase ($g_{\rm ma}<\kappa$).  However, for a sufficiently strong drive field, the above conditions for achieving the $\mathcal{PT}$ symmetry are modified due to the appreciable coupling with the mechanical mode, especially when the strong coupling is reached (Section~\ref{CPA-SC}). Enhanced sideband response~\cite{Huai19}, controllable output field transmission~\cite{Das23}, magnon blockade~\cite{Wang20} and magnon chaos~\cite{Wang19} have been studied in the $\mathcal{PT}$-symmetric(-like) CMM system. In addition, ultralow-threshold phonon lasing is expected to occur in the $\mathcal{PT}$-symmetric CMM system, enlightened by the results in an analogous optomechanical system~\cite{Jing14}.

It is worth noting that for the $\mathcal{PT}$-symmetric CMM system, the stability analysis of the system needs to be given sufficient attention, because the introduction of the gain will strongly modify the stability conditions, making the system more easily tend to be unstable. This becomes particularly important when studying some quantum effects, such as mechanical ground-state cooling and entanglement.

As the counterpart of $\mathcal{PT}$ symmetry, anti-$\mathcal{PT}$ symmetry related phenomena have been demonstrated or predicted in cavity magnonic systems, such as anti-$\mathcal{PT}$-symmetry phase transition~\cite{Zhao20}, enhanced sensing of anharmonicities~\cite{Nair21}, unconventional singularity~\cite{Yang2020} and anti-$\mathcal{PT}$-symmetry enhanced microwave-optics interconversion~\cite{Mukhopadhyay22}, etc. The addition of the mechanical mode by the magnetostrictive effect would make the CMM system a new platform for exploring anti-$\mathcal{PT}$ symmetry and topological properties around EPs.

\begin{figure}[b]
\centering\includegraphics[width=15cm]{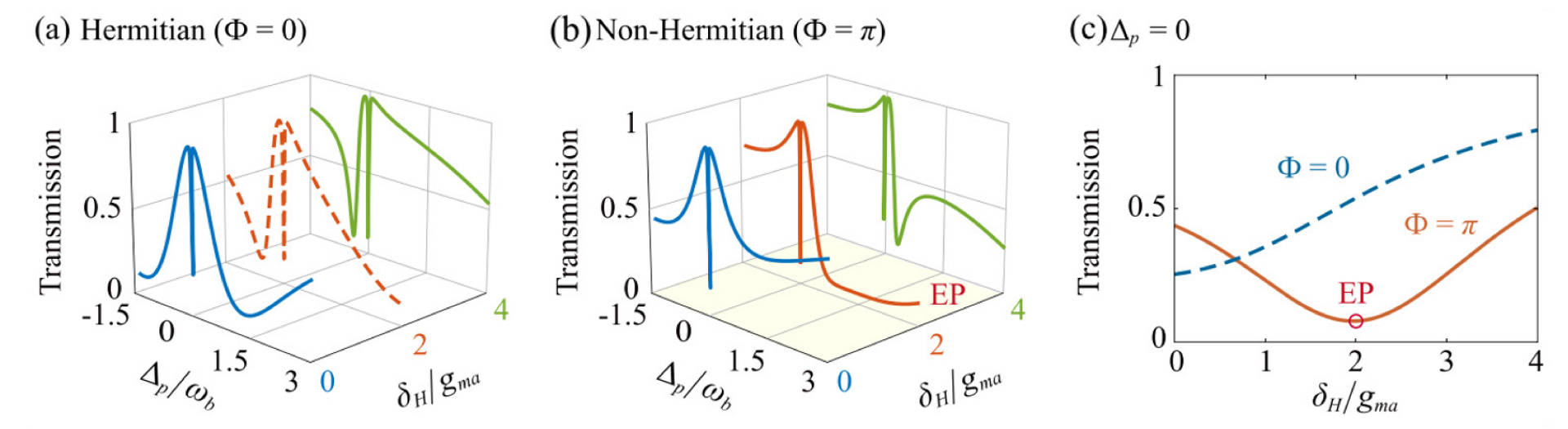} 
\caption{Transmission rate of the probe field versus the magnon-cavity detuning $\delta _{H}=\omega_m-\omega_a$ and the probe-pump detuning $\Delta _{p}= \omega_p -\omega_0 -\omega_b$, with $\omega_p$ ($\omega_0$) being the frequency of the probe (pump) field. The figures are adapted from Ref.~\cite{Lu21}. }
\label{nonH}
\end{figure}

\subsubsection{Exceptional-point-engineered CMM}

Apart from the introduction of the gain into the system as discussed above, the EP can also be achieved in a non-Hermitian CMM system with a dissipative magnon-photon coupling~\cite{Lu21}.  The interaction Hamiltonian of the system includes a dissipative photon-magnon coupling and a coherent magnon-phonon coupling, i.e.,
\begin{equation}
H_{\rm int}/\hbar=g_{\rm ma}(a^{\dagger }m+e^{i\Phi }am^{\dagger })+g_{\rm mb}m^{\dagger}m(b+b^{\dagger }).
\end{equation}
The first term describes the novel non-Hermitian coupling caused by the cavity Lenz effect~\cite{Harder18,Xu19}, and  the coupling phase $\Phi$ describes the competing coherent and dissipative couplings: $\Phi=0$ ($\pi$) for the photon-magnon coherent (dissipative) coupling.

As clearly shown in Figure~\ref{nonH}(a)-(b), an EP emerges due to the presence of the dissipative coupling, which can lead to flexible on-off transmission of the probe field (Figure~\ref{nonH}(c)). It is also shown that slow-to-fast light switching and enhanced mechanical cooling rate can be achieved by operating the system approaching the EP~\cite{Lu21}.  All of these indicate that the EP-assisted CMM system could enable many applications in both classical and quantum regimes.

\subsection{Nonlinear dynamics: self-sustaining dynamics and synchronization}

Most of the studies in {the field of CMM} are based on an approximate linearized magnomechanical interaction under a strong pump. However, the magnomechanical interaction is intrinsically nonlinear.  Including full nonlinearity of the interaction gives rise to rich nonlinear dynamics of the system, such as high-order sidebands (Section~\ref{High-order-sidebands}), self-sustaining dynamics~\cite{WLi23}, and synchronization of mechanical motions~\cite{Cheng23}.  In Ref.~\cite{WLi23}, diverse nonlinear self-sustaining dynamics are predicted, including non-Gaussian phase spreading, amplitude squeezing, and the mixture of multiple limit-cycle states. The nonlinear magnetostriction can also be used to synchronize two mechanical oscillators~\cite{Cheng23}. It shows that by applying a blue-detuned magnon drive, a strong phase correlation can be established between two mechanical oscillators, leading to their synchronization. It also reveals that the strong cavity-magnon linear coupling provides a new degree of freedom to enhance and modulate the synchronization. Apart from the above, many more nonlinear effects remain to be explored, e.g., by exploiting rich nonlinearities of the YIG~\cite{Shen22,YPWang18}.

\subsection{Other magnomechanical platforms}

The preceding sections focus on the CMM system with a dispersive magnetostrictive interaction, typically of a macroscopic ferromagnet (e.g., a sub-mm sized YIG sphere). However, the magnomechanical system can refer to any system that achieves an effective coupling between magnons and phonons, which can be realized in a variety of configurations~\cite{Yuan22,Zhang2023,YLi21}.

\begin{figure}[t]
	\includegraphics[width=1\linewidth]{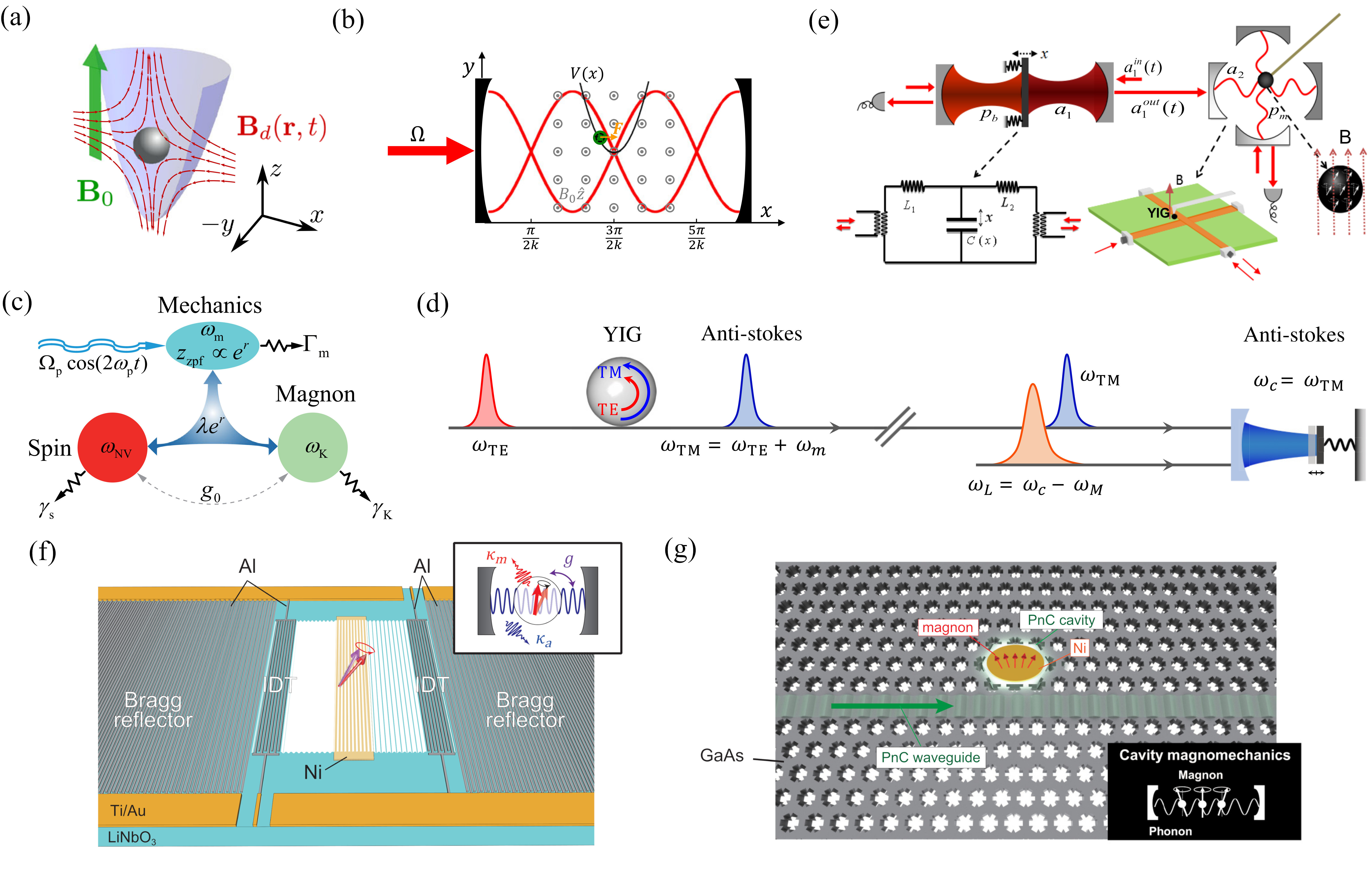}
	\caption{Other magnomechanical configurations. (a)-(c) Hybrid magnomechanical configurations including the CoM motion of a micromagnet for studying acoustic-CoM strong coupling \cite{Gonzalez-Ballestero2020}, ground-state cooling of the CoM motion \cite{Kani2022}, and spin-magnon-phonon strong coupling \cite{Hei2023} and giant enhancement of the magnon-spin coupling \cite{Wang2023}. (d)-(e) Quantum network protocols with magnonic and mechanical nodes connected by optical pulses \cite{LiJ2021} or microwave photons \cite{Tan2021}. (f)-(g) SAW-based CMM using an integrated acoustic cavity \cite{Hatanaka2022} or a phononic crystal cavity \cite{Hatanaka2023}. Figures (a)-(g) are adapted from Refs.~\cite{Gonzalez-Ballestero2020,Kani2022,Hei2023,LiJ2021,Tan2021,Hatanaka2022,Hatanaka2023}, respectively.}
	\label{FIG2}
\end{figure}

Of particular interest is the coupling between the magnonic mode and the center-of-mass (CoM) motion of a spherical micromagnet, e.g., a YIG sphere~\cite{Gonzalez-Ballestero2020,Kani2022,Hei2023,Wang2023}.  The YIG sphere can be trapped in an external harmonic potential (Figure \ref{FIG2}(a)), and an oscillating gradient magnetic field $B_d$ induces the coupling between the magnon mode and the CoM motion via $G \mathrm{cos}(\omega_d t)(m+m^\dagger)(b+b^\dagger)$~\cite{Gonzalez-Ballestero2020}, where $G$ is the coupling rate, $\omega_d$ is the oscillating frequency of the magnetic field, and $b$ denotes the CoM motion. By including also the linear magnetoelastic coupling, strong coupling between the internal acoustic vibration mode and the external low-frequency CoM motion is shown to be possible~\cite{Gonzalez-Ballestero2020}.  By further coupling the levitated YIG sphere to a microwave cavity (Figure \ref{FIG2}(b)), an effective optomechanical-like coupling between the magnon mode and the CoM motion is established, which can be used to cool the CoM motion to its ground state~\cite{Kani2022}. Impressively, this protocol shows the potential to cool the CoM of a sub-centimeter sized sphere to its quantum ground state~\cite{Kani2022}. In addition, by coupling the levitated YIG sphere to a single nitrogen-vacancy spin (Figure \ref{FIG2}(c)), strong tripartite spin-magnon-phonon coupling is promising to be reached via applying a parametric (two-phonon) drive to modulate the CoM motion~\cite{Hei2023}.  The spin-magnon-phonon coupling can also be realized in the configuration of a micromechanical cantilever holding a YIG sphere~\cite{Wang2023}, where the previous CoM motion of the levitated YIG sphere is replaced with the mechanical motion of the cantilever. By applying a parametric drive to squeeze the mechanical motion, both the spin- and magnon-phonon couplings can be exponentially enhanced, leading to giant enhancement of the effective magnon-spin interaction~\cite{Wang2023}.

The magnon and mechanical modes can also establish a long-range coupling by the aid of optical pulses \cite{LiJ2021} or a microwave coaxial cable \cite{Tan2021,Tan2023}. As sketched in Figure~\ref{FIG2}(d), a quantum network protocol consisting of magnonic and mechanical nodes connected by optical pulses is proposed \cite{LiJ2021}. By exploiting both the optomagnonic~\cite{Rameshti22} and optomechanical~\cite{Aspelmeyer14} interactions, it is possible to realize high-fidelity magnon-to-phonon quantum state transfer and entangle two remote magnonic and mechanical nodes~\cite{LiJ2021}. In another configuration combining electromechanical and electromagnonic systems (Figure \ref{FIG2}(e)), a deterministic scheme is offered to establish entanglement and asymmetric steering between a macroscopic mechanical oscillator and a magnon mode~\cite{Tan2021}. 

The phonons can also be the acoustic phonons of the surface acoustic wave (SAW) interacting with ferromagnetic magnons~\cite{Weiler2011,Weiler2012,Dreher2012}. In such a system, the phonon frequency is in the GHz range and nearly resonant with the ferromagnetic resonance, thus possessing a linear magnon-phonon coupling.  In order to enhance the magnon-phonon coupling, an acoustic cavity is designed, formed by either Bragg reflectors \cite{Hatanaka2022} (Figure \ref{FIG2}(f)) or phononic crystal \cite{Hatanaka2023} (Figure \ref{FIG2}(g)), which enables the coherent transduction between magnons and acoustic phonons \cite{Hatanaka2022} and local manipulation of magnons using phonons strongly confined in the phononic crystal cavity~\cite{Hatanaka2023}.

\section*{Data availability statement}
No new data were created or analyzed in this study.

\section*{Acknowledgments}

We thank our colleagues and research teams over the last years for their fruitful and stimulating discussions. This work has been supported by National Key Research and Development Program of China (Grant No. 2022YFA1405200), National Natural Science Foundation of China (Grant Nos. 92265202, 12022507, 12174140) and Scientific Research Foundation for the PhD (Dalian Polytechnic University, No. 7230010115).

\end{document}